\documentclass[twocolumn,superscriptaddress, preprintnumbers , amsmath,amssymb,letterpaper,floatfix]{revtex4}
\usepackage{dcolumn}% Align table columns on decimal point
\usepackage{bm}% bold math
\usepackage{amsmath}
\usepackage{multirow}
\usepackage{graphicx}
\usepackage{float}
\usepackage[colorlinks=true]{hyperref}
\usepackage{color}
\usepackage{fancyhdr}
\begin{document}

 %\twocolumn
 %\baselineskip = 1.5\baselineskip

 \newcommand{\re}{\mathop{\mathrm{Re}}}
 \newcommand{\im}{\mathop{\mathrm{Im}}}
 \newcommand{\D}{\mathop{\mathrm{d}}}
 \newcommand{\I}{\mathop{\mathrm{i}}}
 \newcommand{\E}{\mathop{\mathrm{e}}}
 \newcommand{\unite}[2]{\mbox{$#1\,{\rm #2}$}}
 \newcommand{\myvec}[1]{\mbox{$\overrightarrow{#1}$}}
 \newcommand{\mynor}[1]{\mbox{$\widehat{#1}$}}
 \newcommand{\rmsemit}{\mbox{$\tilde{\varepsilon}$}}
 \newcommand{\exn}{\ensuremath{\epsilon_{xn}}}
 \newcommand{\parmela}{\textsc{parmela}}
 \newcommand{\compton}{\textsc{compton}}
 \newcommand{\um}{\mathrm{\mu m}}
 \newcommand{\wo}{\ensuremath{\mathrm{w_0}}}
\newcommand{\bunits}{\mathrm{photons/(sec\  mm^2\  mrad^2\ 0.1\%)}}
 \newcommand{\zr}{\mathrm{Z_R}}
 \newcommand{\azero}{\mathrm{a_0}}
 \newcommand{\us}{\mathrm{\mu s}}
 \newcommand{\sang}{\ensuremath{\sin \theta}}
 \newcommand{\cang}{\ensuremath{\cos \theta}}
 \newcommand{\bz}{\ensuremath{\beta_z}}
 \newcommand{\bx}{\ensuremath{\beta_x}}
 \newcommand{\ax}{\ensuremath{\alpha_x}}
 \newcommand{\lx}{\ensuremath{\lambda_x}}
 \newcommand{\degree}{\ensuremath{^\circ}}
 \newcommand{\mean}[1]{\mbox{$\langle{#1}\rangle$}}

%\title{Compact X-ray Source using a High Repetition Rate Laser and Copper Linac}
\title{Compact x-ray source based on burst-mode inverse Compton scattering at 100 kHz}
% Force line breaks with \\
\author{W. S. Graves}\affiliation{Massachusetts Institute of Technology, Cambridge, MA 02139, USA}
\author{J. Bessuille}\affiliation{MIT-Bates Laboratory, Middleton, MA 02139, USA}
\author{P. Brown}\affiliation{MIT-Bates Laboratory, Middleton, MA 02139, USA}
\author{S. Carbajo}\affiliation{CFEL, Hamburg, Germany}
\author{V. Dolgashev}\affiliation{SLAC, Stanford, USA}
\author{K.-H. Hong}\affiliation{Massachusetts Institute of Technology, Cambridge, MA 02139, USA}
\author{E. Ihloff}\affiliation{MIT-Bates Laboratory, Middleton, MA 02139, USA}
\author{B. Khaykovich} \affiliation{Massachusetts Institute of Technology, Cambridge, MA 02139, USA}
\author{H. Lin} \affiliation{Massachusetts Institute of Technology, Cambridge, MA 02139, USA}
\author{K. Murari}\affiliation{CFEL, Hamburg, Germany}
\author{E. A. Nanni} \affiliation{Massachusetts Institute of Technology, Cambridge, MA 02139, USA}
%\author{P. Piot} \affiliation{Department of Physics, Northern Illinois University, DeKalb IL 60115,
%USA} \affiliation{Fermi National Accelerator Laboratory, Batavia, IL 60510, USA}
\author{G. Resta} \affiliation{Massachusetts Institute of Technology, Cambridge, MA 02139, USA}
\author{S. Tantawi}\affiliation{SLAC, Stanford, USA}
\author{L.E. Zapata} \affiliation{Massachusetts Institute of Technology, Cambridge, MA 02139, USA}
\affiliation{CFEL, Hamburg, Germany}
\author{F.X. K{\"a}rtner}\affiliation{Massachusetts Institute of Technology, Cambridge, MA 02139, USA}
\affiliation{CFEL, Hamburg, Germany}
\author{D.E. Moncton}\affiliation{Massachusetts Institute of Technology, Cambridge, MA 02139, USA}
\date{\today}% It is always \today, today,

\begin{abstract}
A design for a compact x-ray light source (CXLS) with flux and brilliance orders of magnitude beyond existing laboratory scale sources is presented.  The source is based on inverse Compton scattering of a high brightness electron bunch on a picosecond laser pulse.  The accelerator is a novel high-efficiency standing-wave linac and RF photoinjector powered by a single ultrastable RF transmitter at x-band RF frequency.  The high efficiency permits operation at repetition rates up to 1~kHz, which is further boosted to 100~kHz by operating with trains of 100 bunches of 100~pC charge, each separated by 5~ns.  The entire accelerator is approximately 1 meter long and produces hard x-rays tunable over a wide range of photon energies.  The colliding laser is a Yb:YAG solid-state amplifier producing 1030~nm, 100~mJ pulses at  the same 1~kHz repetition rate as the accelerator.  The laser pulse is frequency-doubled and stored for many passes in a ringdown cavity to match the linac pulse structure.  At a photon energy of 12.4~keV, the predicted x-ray flux is $5 \times 10^{11}$ photons/second in a 5\% bandwidth and the brilliance is $2 \times 10^{12}\  \bunits$ in pulses with RMS pulse length of 490~fs.  The nominal electron beam parameters are 18~MeV kinetic energy, 10~microamp average current, 0.5~microsecond macropulse length, resulting in average electron beam power of 180 W.  Optimization of the x-ray output is presented along with design of the accelerator, laser, and x-ray optic components that are specific to the particular characteristics of the Compton scattered x-ray pulses.
\end{abstract}

\maketitle
\thispagestyle{fancy} % Comment this line out if don't want header & footer.

%%%%%%%%%%%%%%%%%%%%%%%%%%%%%%%%%%%%%%%%%%%%%%%%%%%%%%%%%%%%%%%%%%%%%%%%%%

\section{Introduction}
The main rationale for new x-ray sources is that since their discovery in 1895, x-rays have been the single most powerful technique for determining the structure of all forms of condensed matter. Through increasingly powerful imaging, diffraction, and spectroscopic techniques, physicists, chemists, biologists, and medical doctors, as well as quality-control inspectors, airline passenger screeners, and forensic scientists have resolved the structural detail and elemental constituency on length scales from inter-atomic spacing to the size of the human body. Every day, that knowledge underpins our modern technologies, our health, and our safety. Quite remarkably, for the first 70 of these 119 years, x-ray sources changed little from the original tube~\cite{roentgen_1896} used by R\"{o}ntgen in the discovery of x-rays. Even today, the x-ray technology used in universities, industrial labs and hospitals derives from this primitive electron tube with incremental improvements. 

Today, however, the benchmark for x-ray performance is set by large accelerator-based synchrotron radiation facilities, of which more than 60 exist worldwide. Interestingly, despite this large investment in major facilities, 15 of the 19 Nobel Prizes awarded for x-ray-based discoveries used conventional or rotating-anode-based x-ray sources. The wide availability of these modest sources, their ease of use, and their capability to test new ideas without the barriers of schedule, travel, and expense of the major facilities has led to a remarkable array of scientific breakthroughs. In the present work we aim to exploit recent advances in accelerator and laser technologies to produce a CXLS thats offer the potential for a more powerful x-ray source of appropriate size, complexity, and cost for modern laboratories and hospitals.

X-ray beam performance from a CXLS would not yet approach the performance of the mega-facilities in general, but in critical parameters such as pulse length and source size the compact sources would surpass them. The result would be major advances for example in ultrafast x-ray science and diagnostic phase-contrast imaging. For ultra-fast x-ray diffraction, to understand matter far from equilibrium or study chemical reactions in real time, CXLS technology can produce powerful pulses of x-rays having time duration well under a picosecond. New phase-contrast imaging methods, already demonstrated on large synchrotrons, could be available in hospitals to provide a new tool in the battle against breast cancer or heart disease.

\begin{figure*}[ht]
\includegraphics[width=0.90\textwidth]{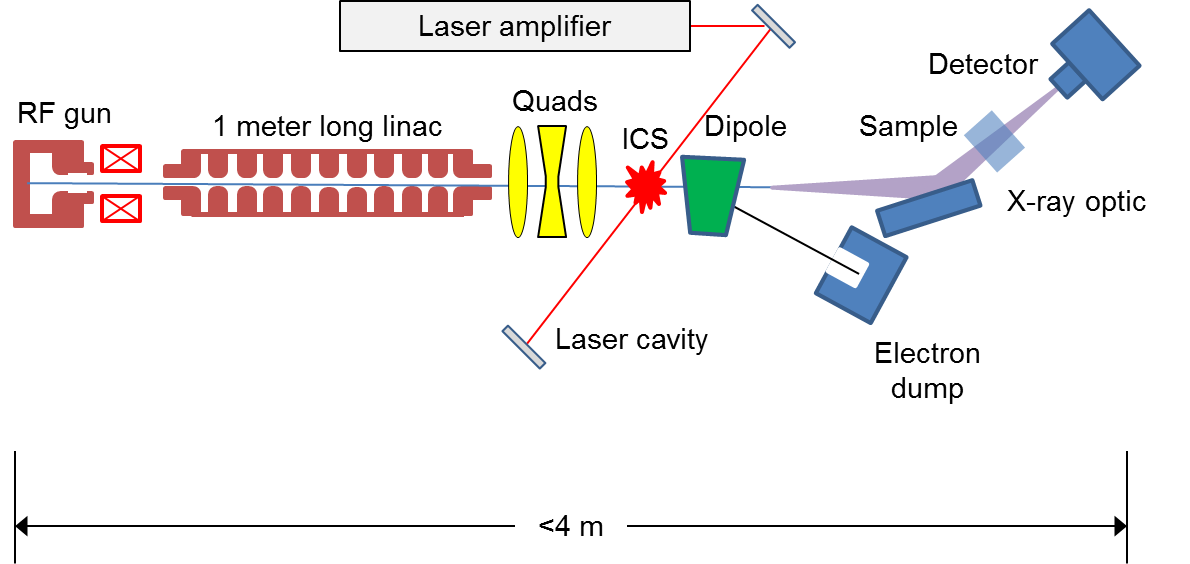}
\caption{Accelerator, laser, and x-ray components needed for the ICS x-ray source.  Total length from RF gun cathode to ICS interaction point is 2.5~m.  Length of x-ray beamline depends on experiment but is typically of order 1~m.  Relatively large x-ray divergence of several mrad leads to shorter beamlines.}
\label{fig:layout}
\end{figure*}

In addition to entirely new applications, compact sources offer many other advantages. The pace of drug discovery could be accelerated by providing structure solutions immediately upon crystallization, rather than waiting precious days or weeks for access to a synchrotron. Universities will have the ability to train students in these powerful emergent applications of x-rays without the often prohibitive constraints of travel and the limitations of beam time. Electronic chip manufacturing facilities could perform in-situ metrology of today's three dimensional nanometer sized structures.  Museums and other cultural institutions could perform in-house x-ray analysis on historic works of art.  Advanced major facilities such as the National High Magnetic Field Laboratory and Spallation Neutron Source at ORNL can benefit from a compact source to combine x-ray analysis with magnetic and neutron studies.  Furthermore, the characteristics of the x-rays produced by such compact sources can be improved through research and development, increasing flux and coherence with the ultimate goal of achieving a compact x-ray source with fully coherent \cite{graves_2012, graves_2013} hard x-ray beams.

Inverse Compton scattering is the upconversion of a low energy laser photon to a high energy x-ray by scattering from a relativistic electron.  Figure \ref{fig:layout} shows the layout and geometry of the interaction with a near head-on collision between the laser and electron beams.  The scattered x-rays emerge in the same direction as the electrons.  The physics of ICS is nearly identical to spontaneous synchrotron emission in a static magnetic undulator as used at the large traditional synchrotron facilities, but because the wavelength of the laser is much shorter than a static undulator period, the energy required of the electrons to make hard x-rays is orders of magnitude lower than the large synchrotrons, reducing the size and cost of the compact ICS source by orders of magnitude.

\begin{figure*}[ht]
\includegraphics[width=0.90\textwidth]{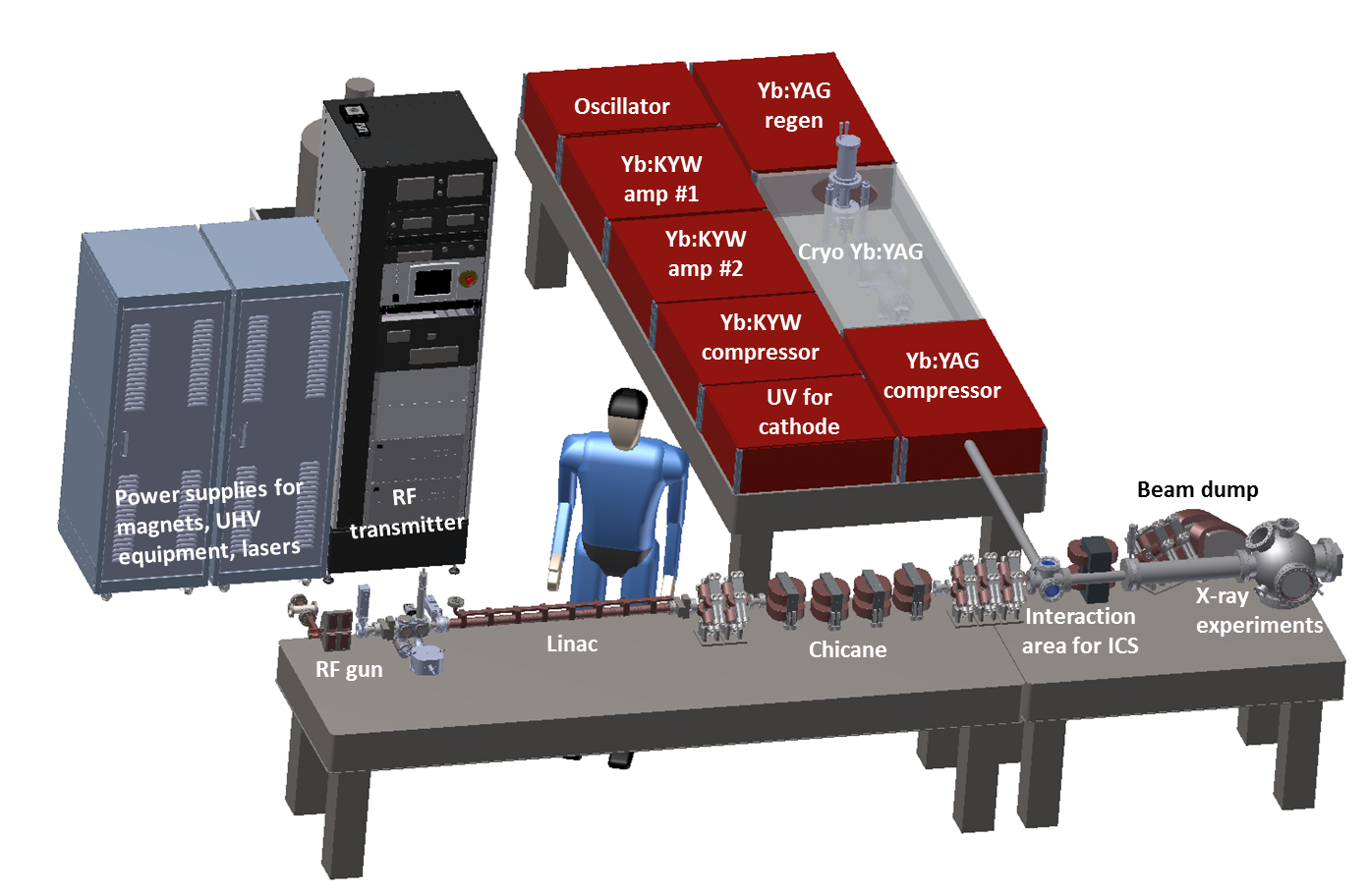}
\caption{CAD layout of the components for the compact x-ray source showing the lasers including Yb:KYW that produces electrons via photoemission and the cryo Yb:YAG amplifier used for ICS.  Accelerator components shown include the RF gun, short linac and transport magnets.  The cabinets house the RF transmitter and power supplies for magnets, vacuum equipment, and lasers.}
\label{fig:cadICSassembly}
\end{figure*}

Linac-based Compton-scattering x-ray sources have shown promising results at low repetition rates \cite{brown_2004a}. However, past linac-based ICS x-ray experiments depended on accelerators and lasers built for other purposes such as injection into a synchrotron that had modest performance requirements, and produced x-ray beams of relatively low brilliance due to the lack of specific engineering to ICS requirements.  Recent developments in laser and accelerator technology, along with a focused effort to design a dedicated ICS x-ray source promise to significantly improve the source properties and to open up new scientific frontiers, particularly in ultrafast dynamics and spectroscopy.  In this work we focus on the properties of a linac-based CXLS in preference to ring-based output.  The main advantages of compact rings (see \cite{xu_2014} for a good discussion) are their high repetition rate, high average current relative to a linac, and historically better stability.  However the dynamics in rings result in relatively long electron bunches with large emittance; both of these effects are detrimental to x-ray production with the result that a relatively low repetition rate linac can produce the same or greater flux than a ring.  The pulse format for a linac is flexible and determined by the photocathode laser.  The output pulses are subpicosecond without compression and substantially shorter than that with compression.  The electron beam phase space may be manipulated in sophisticated ways, e.g. emission of arrays of electrons from nanotips or diffraction crystals, or filtering for transverse momentum or energy.  There is great flexibility with a single pass device where the electron bunches do not relax into an equilibrium distribution.  Rings are substantially larger and more complex than a linac; they require a full energy linac for injection, and a high-finesse, high power laser cavity as well as all of the ring equipment.  We think it will eventually be possible to produce coherent x-rays via ICS, but this will require the best possible electron beam which will only be produced by a linac.  There is a perception that linacs have significant jitter in timing, energy, and beam position.  However that view derives from linacs that were designed for injection into synchrotrons and required only modest stability.  Modern solid-state high power RF equipment and precision magnet power supplies have stabilized operation to the point where femtosecond jitter and micron position stability are attainable.

Although the baseline CXLS performance will not match the standard of large 3rd and 4th generation facilities, CXLS sources are expected to provide x-ray parameters similar to that of bending magnet synchrotron beam lines for a small fraction of the cost, far beyond existing lab source performance, and in some parameters such as pulse length and source size, will significantly exceed what is possible at the major facilities.  The current situation in x-ray science is that experiments are either done at the large facilities where the beam brightness is about $10^{21}\ \bunits$, or in home labs where the best rotating anodes \cite{rigaku_frx} have a brightness of $10^9$ and flux of $6 \times 10^9$ per second.  That enormous gap - 12 orders of magnitude in brilliance - is similar to the gap in capability between an abacus and a supercomputer.  The CXLS is equivalent to a laptop, and like the laptop it is closer to a supercomputer than an abacus.  It's qualitative and quantitative impact on x-ray science is likely to be enormous, in ways that are difficult to foresee because a source with this brilliance, size, and cost does not exist today.

\begin{figure*}[ht]
\includegraphics[width=0.90\textwidth]{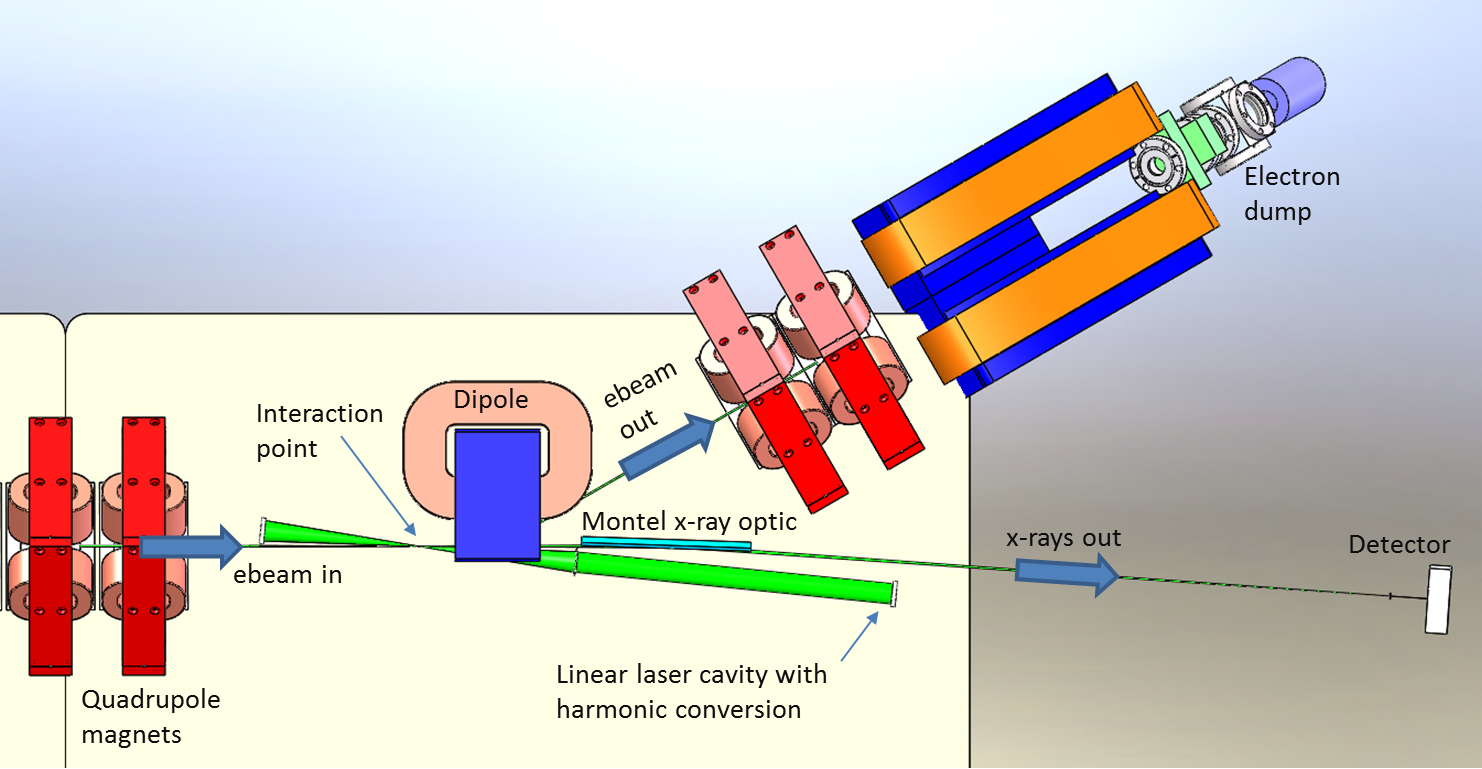}
\caption{Laser, electron, and x-ray components near the interaction point are intricately arranged.  Each of the three beams needs a strong lens near the IP that does not interfere with the other beams.  The electron beam crosses the laser beam at an angle of 50~mrad so that the x-rays avoid the laser optic.  The electrons bend out of the way immediately after the collision at the IP.  An x-ray optic is installed 20~cm downstream of the IP to collect and focus or collimate the divergent x-ray beam.}
\label{fig:IP01}
\end{figure*}

%%%%%%%%%%%%%%%%%%%%%%%%%%%%%%%%%%%%%%%%%%%%%%%%%%%%%%%%%%%%%%%%%%%%%%%%%%

\section{Compact Source Components}

A CAD drawing of the major components of the compact source is shown in Figure \ref{fig:cadICSassembly} including the electron accelerator, laser amplifiers and laser ringdown cavity inside which x-ray production takes place, x-ray optics, and RF transmitter that supplies power to the accelerator.  For a compact source it is important to design the entire system to be small, and not to overlook ancillary items such as power supplies, chillers or cryogenic equipment.  The ancillary equipment for compact sources can dominate the size and cost of the system.  However the advent of high-power solid-state power sources in recent years for both lasers and accelerators has dramatically reduced the overall footprint of the equipment while improving stability and efficiency.  The system shown requires only three standard equipment racks for all of its support equipment.

Compact sources are meant to be installed in scientific, medical, and industrial labs that generally do not have accomodation for sources of ionizing radiation requiring retrofit installation of shielding.  Electrons at energies above 10~MeV produce neutrons requiring bulky and heavy shielding, and so the accelerator is typically installed in a shielded vault or room, which is a significant impediment to widespread use of these powerful x-ray sources.  However, the compact source shown in Figure \ref{fig:cadICSassembly} is small enough (0.3 $\times$ 3~m) to have local shielding fitted directly to the device enclosing just the accelerator components in concrete blocks.

A detailed view of the laser-electron interaction area is shown in Figure \ref{fig:IP01}.  Each of the three beams - laser, electron, and x-ray - are present at the very small interaction point (IP), thus as for any beam brought to a tight focus, a lens is required nearby with a large numerical aperture.  One of the main challenges for the source designer is to fit these diverse lens types into the limited space available.  In the figure, the electron beam enters from the left and is focused by a quadrupole triplet lens (last quadrupole is visible) to a $3\ \um$ RMS spot at the IP.  Immediately after the collision with the laser the electrons of kinetic energy 18~MeV are bent by a dipole and transported to a beam dump by the quadrupoles visible at the upper right of the figure.  The electrons arrive at the IP as a bunch train, 100 bunches (each about 1 ps long) arrive in a $0.5\ \us$ long train.  The train repeats at a 1 kHz repetition rate providing an effective repetition rate of 100~kHz.  A single 50~mJ, picosecond laser pulse is coupled into a 4f confocal linear cavity.  The laser beam is the green beam reaching a focus between a mirror and lens in the figure.  The laser also operates at 1 kHz and each pulse rings down inside the cavity with a roundtrip time of 5~ns to match the 100 bunch electron train.  The multiple electron bunches all interact with the same decaying laser pulse.  The laser cavity is set at a 50~mrad angle with respect to the electron beam to allow the electrons and x-rays to miss the laser optics.  The x-rays propagate in the same direction as the electrons.  The laser direction does not impact the x-ray direction although the small collision angle does cause a minor reduction in flux and a small shift in x-ray wavelength relative to a head-on collision.  The opening angle of the x-rays is relatively large at approximately 10~mrad so that a nearby x-ray optic is required to collect the flux.  The optic is shown just downstream of the IP between the laser and electron beams.

%%%%%%%%%%%%%%%%%%%%%%%%%%%%%%%%%%%%%%%%%%%%%%%%%%%%%%%%%%%%%%%%%%%%%%%%%%

\begin{figure*}[ht]
\includegraphics[width=0.90\textwidth]{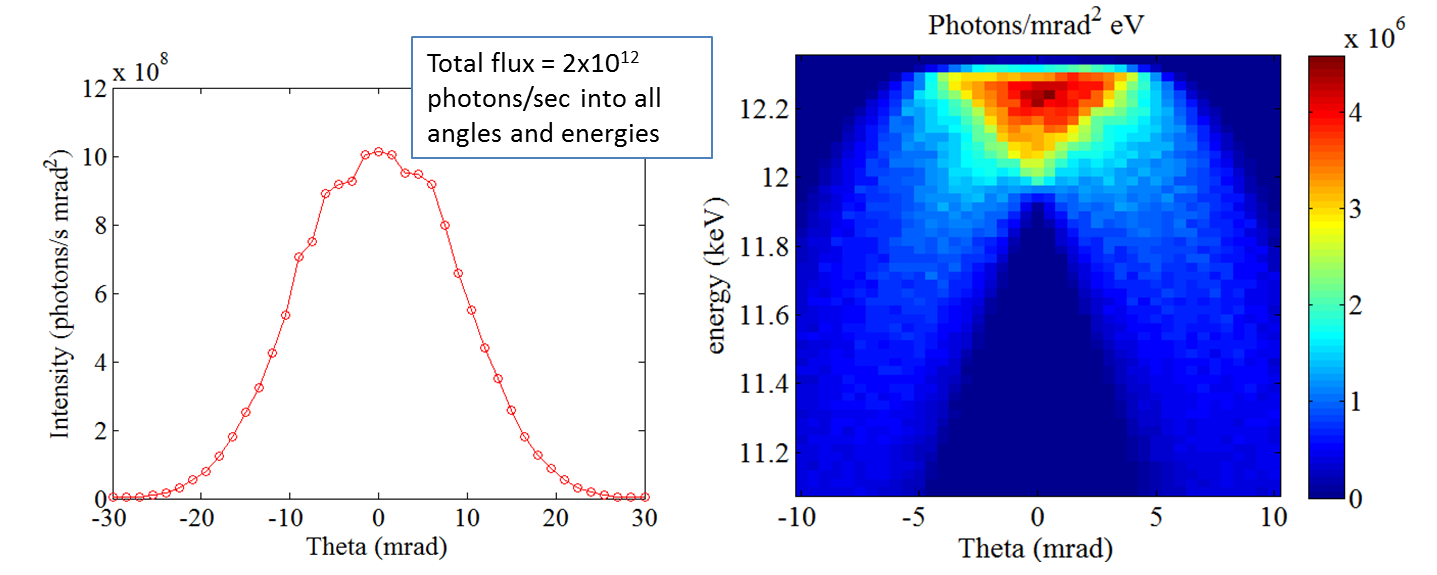}
\caption{Plots of total x-ray output near 12 keV.  The flux at 100 kHz is $2\times10^{12}$ photons/sec into all angles and energies.  Left plot shows flux vs angle and right plot shows color coded intensity vs angle and photon energy.  The on-axis bandwidth is represented by the vertical width of the higher intensity area at $\theta = 0$, and is narrower than the total bandwidth over all angles.  The off-axis photons are lower energy and wider bandwidth than on-axis emission.}
\label{fig:xraytotal12keV}
\end{figure*}

\section{X-ray source optimization}

For a scattering process like ICS, the highest flux is produced by creating the densest target in order to increase the probability of scattering.  High density is achieved by squeezing the electron and laser beams in each of their dimensions.  Thus the laser pulse is short in time and is focused to a small waist.  Likewise the electron beam is focused to a small spot and must also be short in time to interact efficiently with the laser.  The x-ray source is optimized to have a small, radially symmetric size of a few microns and to have an opening angle of a few mrad.  This source contrasts with a typical synchrotron beam that has a source size of $100\ \um$ and an opening angle of perhaps 100~$\mu$rad.  However, because the product of source size and divergence of the CXLS is similar to a synchrotron beam, optics can be used to collimate the ICS beam resulting in a larger virtual source with a smaller opening angle similar to synchrotron radiation.

There are of course limits to improvements in x-ray output by focusing to ever smaller spots.  For the electron beam, the emittance is the determining factor.  The emittance of an electron beam at a focus is just the product of the beam size and its divergence.  Making a smaller focus means that the divergence is larger, and eventually the spread in angles of the electron trajectories will become the dominant cause of increasing bandwidth and opening angle of the x-rays, lowering the beam brilliance.  The laser focus size should be similar to the electron beam size to maximize the interaction, but producing a small laser focus also sets limits on acceptable pulse lengths.

The optimum pulse lengths for both the electron and laser beams depend on the transverse focus size.  In the regime where we will operate the laser intensity is relatively low so that nonlinear effects \cite{krafft_2004, hartemann_2005b, albert_2011} are weak and x-ray production scales as the square of the laser intensity.  In this case it is important for the electron-laser interaction to take place within the Rayleigh diffraction length $\zr = \pi \wo^2/\lambda_L$ of the laser beam where $\wo$ is the $1/e^2$ laser waist radius.  $\zr$ can become much shorter than the laser pulse length for small \wo, reducing the x-ray flux.  The shortest laser pulse length is generally determined by the laser bandwidth, which depends on the particular laser material chosen.  The laser needs to be able to produce short pulses with high pulse energy and high average power, limiting the selection of materials to a few candidates.   The electron bunch is also constrained to interact within $\zr$ so it must be short as well.  Note that the output x-ray pulse length depends only on the electron bunch length and not the laser length.  This is the same dynamic as in undulator radiation where the x-ray pulse length does not depend on undulator length.  The quantitative effects of pulse length and focus size are shown in numerical studies below.  First we derive relations that guide the design of the electron and laser beams, and resulting x-ray source.

The resonant wavelength for ICS is~\cite{ride_1995}
\begin{equation} \label{eq:resonance}
\lambda_x = \frac{\lambda_L}{2 \gamma^2 (1-\beta_z \cos\phi)} \left(1+ \frac{\azero^2}{2} + \gamma^2 \theta^2\right)
\end{equation}
where $\lambda_L$ is the laser wavelength, $\gamma$ is the electron energy in units of rest mass, $\phi$ is the electron-laser collision angle ($\phi = \pi$ for head-on), and $\azero = \frac{e E \lambda_L}{2 \pi mc^2}$ is the dimensionless vector potential of the laser field that plays the same role in ICS as the undulator parameter $\mathrm{K}$ \cite{kim} does for undulator radiation.  The angle between the electron direction of motion and an observer is $\theta$.  For low laser intensity and on-axis emission from a head-on collision with a relativistic beam the resonant wavelength is $\lambda_{x0} = \frac{\lambda_L}{4 \gamma^2}$. For high laser intensity with multi Joule pulses, $\azero \ge 1$, harmonic power and distortion of the fundamental linewidth become important.  However, in the present case where high repetition rate is important the peak intensity is modest, $\azero \le 0.1$, and a weakly nonlinear approximation is used to accurately model the on-axis spectrum.

\begin{figure*}[ht]
\includegraphics[width=0.80\textwidth]{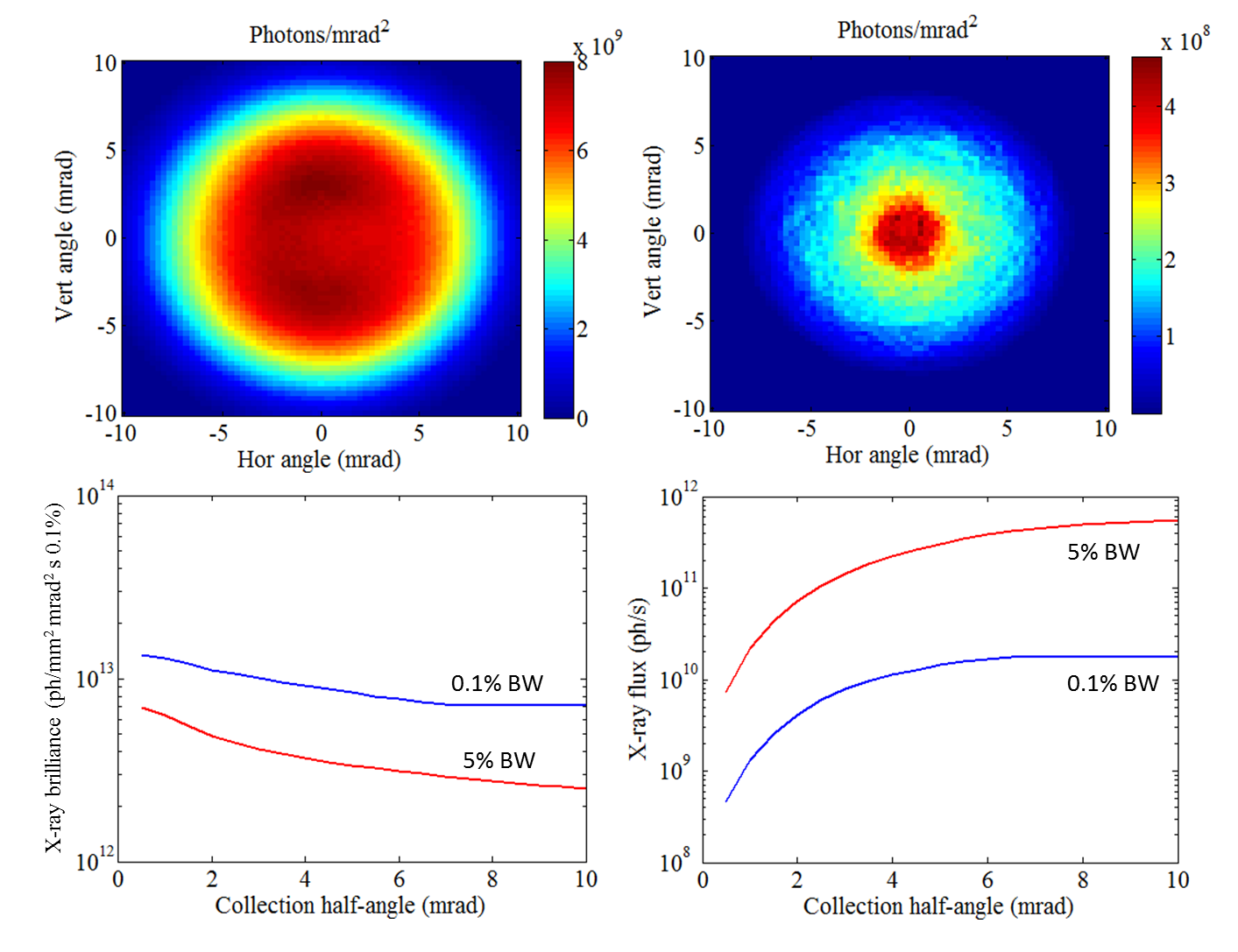}
\caption{Upper plots show x-ray intensity vs opening angle for 5\% bandwidth (left) and 0.1\% bandwidth (right) at 12~keV.  Lower plots show brilliance (left) and flux (right) vs collection angle.  The narrow bandwidth radiation has higher brilliance but lower flux.  Flux for both cases is contained within a half angle of 5~mrad.}
\label{fig:fluxbrilliance}
\end{figure*}

The x-ray bandwidth is determined by several factors including the laser bandwidth, which represents the minimum possible x-ray bandwidth, the electron beam energy spread and emittance, and the wavelength shift represented by the $\azero$ factor in Eq. \ref{eq:resonance} that depends on the time-varying laser intensity.  Figure \ref{fig:xraytotal12keV} shows the complex interplay of bandwidth, central energy, and collection angle.  The minimum x-ray bandwidth is the inverse of the number of laser periods $1/N_L$ similar to the relationship between undulator radiation bandwidth and number of undulator periods.  As an example, a 1~ps laser at 515~nm wavelength has 580 periods and so the minimum bandwidth is about 0.2\%.  This value may be broadened by several effects.  From equation \ref{eq:resonance} the contribution of electron energy spread to broadening is
\begin{equation}
\left(\frac{\Delta \lambda_x}{\lambda_x}\right)_{\text{energy spread}} = 2 \frac{\Delta \gamma}{\gamma}.
\end{equation}
From numerical simulation the expected electron energy spread is 0.1\% in which case the broadening is similar to the fundamental line width.  Another effect is the intensity-dependent term $\azero^2/2$ in equation \ref{eq:resonance}, which results in a red shift of the emission due to a slowing down of the average electron z-velocity in strong laser fields.  A conservative estimate of the bandwidth due to this effect is
\begin{equation}
\left(\frac{\Delta \lambda_x}{\lambda_x}\right)_{\text{laser intensity}} = \frac{\azero^2}{2}.
\end{equation}

For the present case the maximum $\azero = 0.1$ at the focus so that the relative broadening is at the 1\% level.  This estimate is conservative because the probability of emission also scales as $\azero^2$ so that most of the emission occurs near the maximum value of $\azero$.  The electron emittance affects the bandwidth through the variation in electron-laser collision angle $\phi$ and the change in apparent observation angle $\theta$.  At a focus, the normalized electron emittance is $\exn = \gamma \sigma_x \sigma_{x'}$ where $\sigma_x$ is the RMS beam size and $\sigma_{x'}$ is the RMS spread in electron angles relative to the axis.  Equating $\sigma_{x'}$ to the effective observation angle $\theta$ in equation \ref{eq:resonance} results in a broadening of 
\begin{equation}
\left(\frac{\Delta \lambda_x}{\lambda_x}\right)_{\text{emittance}} = \gamma^2 \theta^2 = \frac{\exn^2}{\sigma_x^2}.
\end{equation}
As with the other contributions, we wish to limit this effect to $<1\%$ so that the electron focus size $\sigma_x \ge 10 \exn$.  In the present case the emittance is $0.2\times 10^{-6}$ m-rad so that the electron focus size should be $\sigma_x \ge 2 \um$.

Turning to numerical simulations to study these effects, we run several codes to create detailed time-dependent models of the electron and laser beams and the x-ray beams they produce.  The electron beam is modeled starting from photoemission at the cathode through acceleration and transport to the interaction point (IP) with the code \parmela \cite{parmela}, a time-dependent PIC code including space charge effects.  The laser is modeled with TracePro and Oxalis-Laser codes.  The electron and laser simulations are described in more detail in the technical sections that follow.  The resulting laser and electron pulses are then input into the code \compton \cite{brown_2004b} that performs time-dependent 3D simulations of the incoherent ICS process including weakly nonlinear effects.  This code has been benchmarked \cite{brown_2004a} against ICS experiments.

\begin{figure*}[ht]
\includegraphics[width=0.90\textwidth]{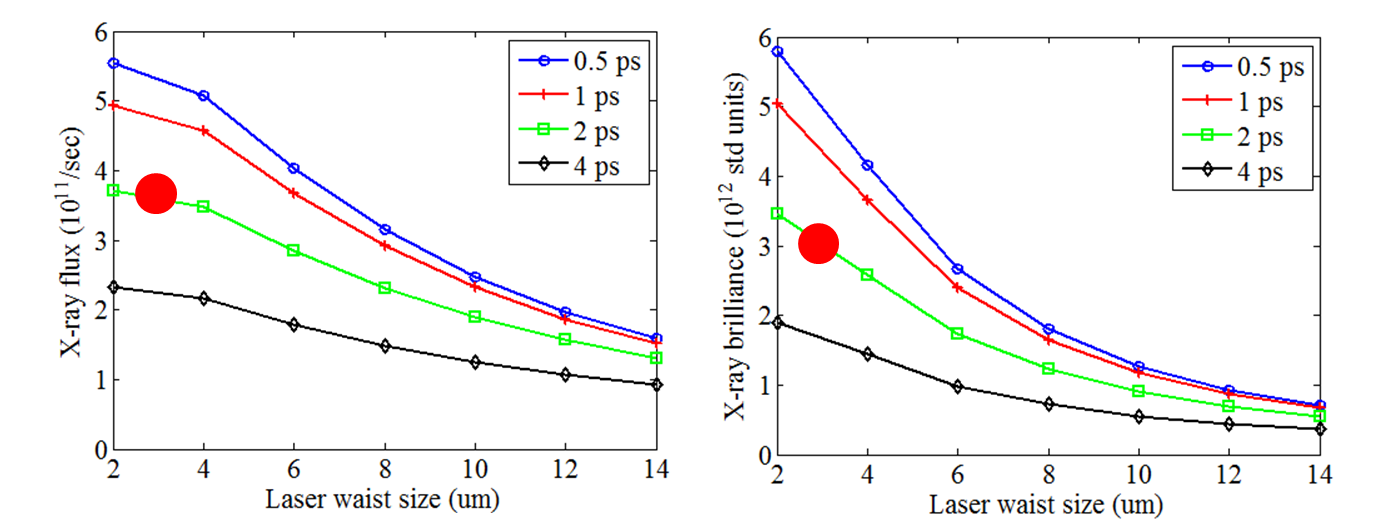}
\caption{Plots of flux (left) and brilliance (right) as a function of $1/e^2$ laser focus size \wo.  Each curve represents a different laser FWHM pulse length from 0.5 to 4~ps.  The red circle in each plot marks the design operating point, which is an optimization of laser gain and bandwidth, average power, and peak power.}
\label{fig:fluxvpulselength}
\end{figure*}

Figure \ref{fig:xraytotal12keV} shows output from \compton, with plots of total flux vs angle and a color plot of x-ray intensity vs angle and photon energy.  Inspecting the righthand plot, the vertical width of the intensity band represents energy bandwidth.  It is apparent that the on-axis bandwidth is quite narrow (0.7\% in this case due to all of the effects described above) but that as radiation is collected at larger angles the bandwidth is substantially broadened and emitted at lower photon energy.  These off-axis effects are due to the $\gamma^2 \theta^2$ term in equation \ref{eq:resonance} creating an energy-angle correlation.  The same phenomenon occurs with undulator radiation; however, the low electron energy used in ICS (one of its primary advantages) results in a larger opening angle of the radiation.  Note that even though the color plot is not as intense at larger angle, the integrated flux at those angles is substantial and useful for experiments that can tolerate the wider bandwidth.  The radiation contained within the central on-axis bandwidth is emitted into a fairly narrow cone of 5~mrad half angle.  

The upper plots of Figure \ref{fig:fluxbrilliance} show x-ray intensity vs horizontal and vertical angles for bandwidth windows of 5\% and 0.1\%.  The acceptable bandwidth depends on the application, with e.g. Laue scattering and SAXS techniques able to use a broad bandwidth, while macromolecular crystallography and other techniques require narrow bandwidth.  For the set of parameters under study the average flux into 5\% bandwidth is $5\times10^{11}$~photons/sec while for a narrow 0.1\% window it is $3\times10^{10}$~photons/sec.  The lower plots of Figure \ref{fig:fluxbrilliance} show how the brilliance and flux scale with collection angle for narrow and wide bandwidth.

Assuming Gaussian laser and electron beam profiles, an analytic expression for the total x-ray dose produced by a head-on inverse Compton  scattering interaction is given by \cite{brown_2004c}
\begin{equation} \label{eq:flux}
N_x = \frac{N_e N_{L} \sigma_T}{2 \pi \left(\sigma_L^2+ \sigma_x^2\right)} FF.
\end{equation}
In equation \ref{eq:flux} $\sigma_T$ is the total Thomson cross section, $N_e$ is the total number of electrons, and $N_{L}$ is the total number of photons in the laser beam.  The term $FF$ is a form factor, less than unity, that depends on RMS pulse durations $\Delta t_L$ and $\Delta t_e$, and beam spot sizes $\sigma_L$ and $\sigma_x$ at the interaction point for the laser and electron beams.  It represents the degradation of the interaction efficiency for cases where the pulse durations exceed the interaction diffraction lengths of the laser and electron beams.  The resulting x-ray brilliance at a collision repetition rate of $F_{rep}$ is
\begin{equation} \label{eq:brilliance}
B_x \approx 1.5 \times 10^{-3} \frac{N_e N_L \sigma_T \gamma^2}{(2 \pi)^3 \exn^2 \sigma_L^2} F_{rep}.
\end{equation}

\begin{figure}[ht]
\includegraphics[width=0.45\textwidth]{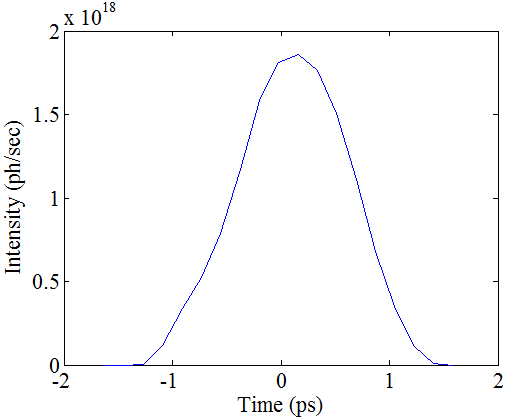}
\caption{Plot of the simulated time duration of the x-ray pulse at 12~keV.  The time duration is nearly equal to the electron bunch length and does not depend on the laser pulse length.  Figure shown is without compression: electron bunch compression would produce significantly shorter pulses.}
\label{fig:xrayvstime}
\end{figure}

\begin{figure*}[ht]
\includegraphics[width=0.90\textwidth]{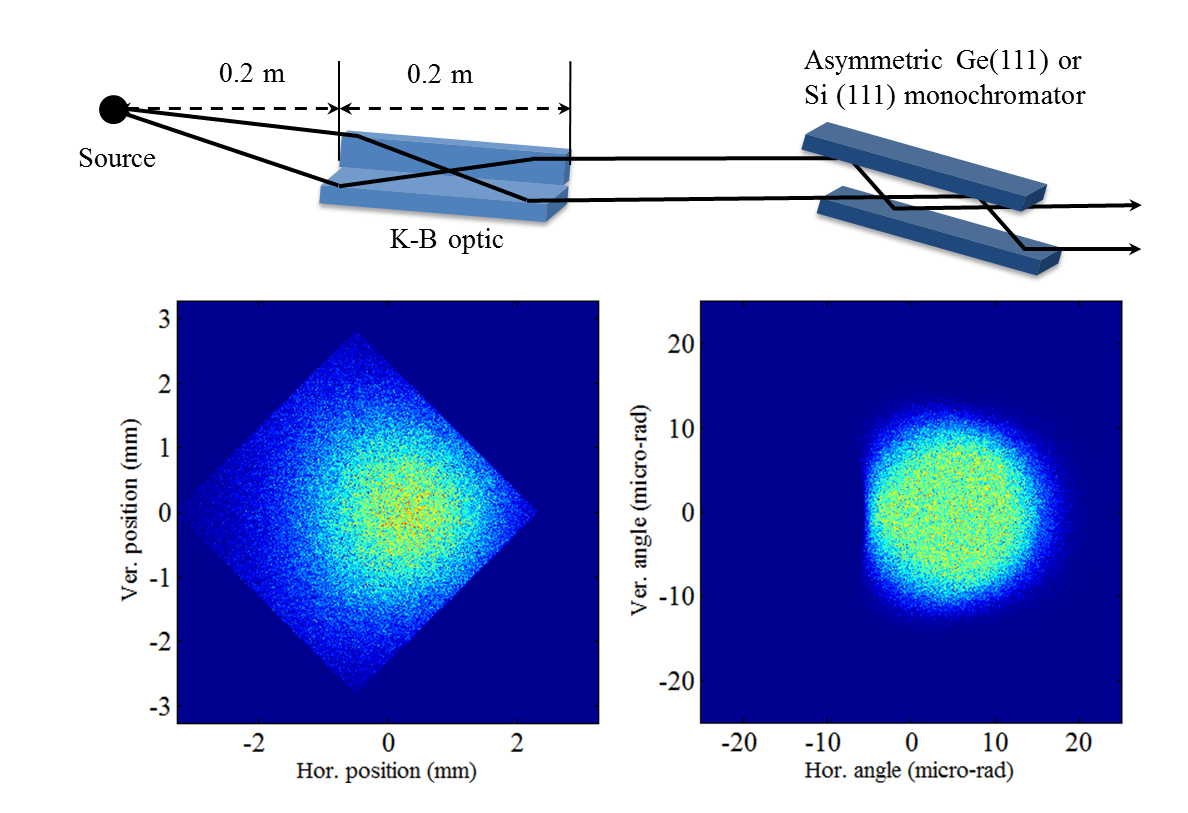}
\caption{Geometry and simulation results for collimating optics.  A nested KB mirror with graded multilayer coating collects 5\% bandwidth radiation at 24~mrad grazing angle.  The ICS source size ($2\ \um$) and opening angle (4~mrad) are converted to a synchrotron-like beam with output RMS beam size of 0.94~mm and RMS divergence of $5.6\ \mathrm{\mu rad}$.  Total efficiency is 40\% including 80\% collection efficiency and 70\% surface reflectivity.}
\label{fig:collimation}
\end{figure*}

\begin{table*}[ht]
\caption{Estimated performance at 0.1\% and 5\% bandwidth for 12.4~keV x-rays from the compact source.}
\label{tab:xray}
\begin{center}
\begin{tabular}{l c c c c}\hline\hline\
Parameter & 0.1\% bandwidth & 5\% bandwidth  &  Units \\
\hline
Photon energy & 12.4  & 12.4 & keV  \\
Average flux  &  $2\times 10^{10}$ & $5\times 10^{11}$   & phot/s  \\
Average brilliance  &  $7\times10^{12}$ & $2\times10^{12}$   & $\bunits$ \\
Peak brilliance & $3\times10^{19}$ & $9\times 10^{18}$ & $\bunits$ \\
RMS hor. opening angle & 3.3 & 4.3 & mrad\\
RMS ver. opening angle & 3.3 & 4.3 & mrad\\
RMS hor. source size & 2.4 & 2.5 & $\um$ \\
RMS ver. source size & 1.8 & 1.9 & $\um$\\
RMS pulse length & 490 & 490 & fs  \\
Photons/pulse  &  $2\times 10^5$ & $5\times 10^6$   & --  \\
Repetition rate & 100 & 100 & kHz \\
\hline \hline
\end{tabular}
\end{center}
\vspace{-3mm}
\end{table*}

\begin{figure*}[ht]
\includegraphics[width=0.90\textwidth]{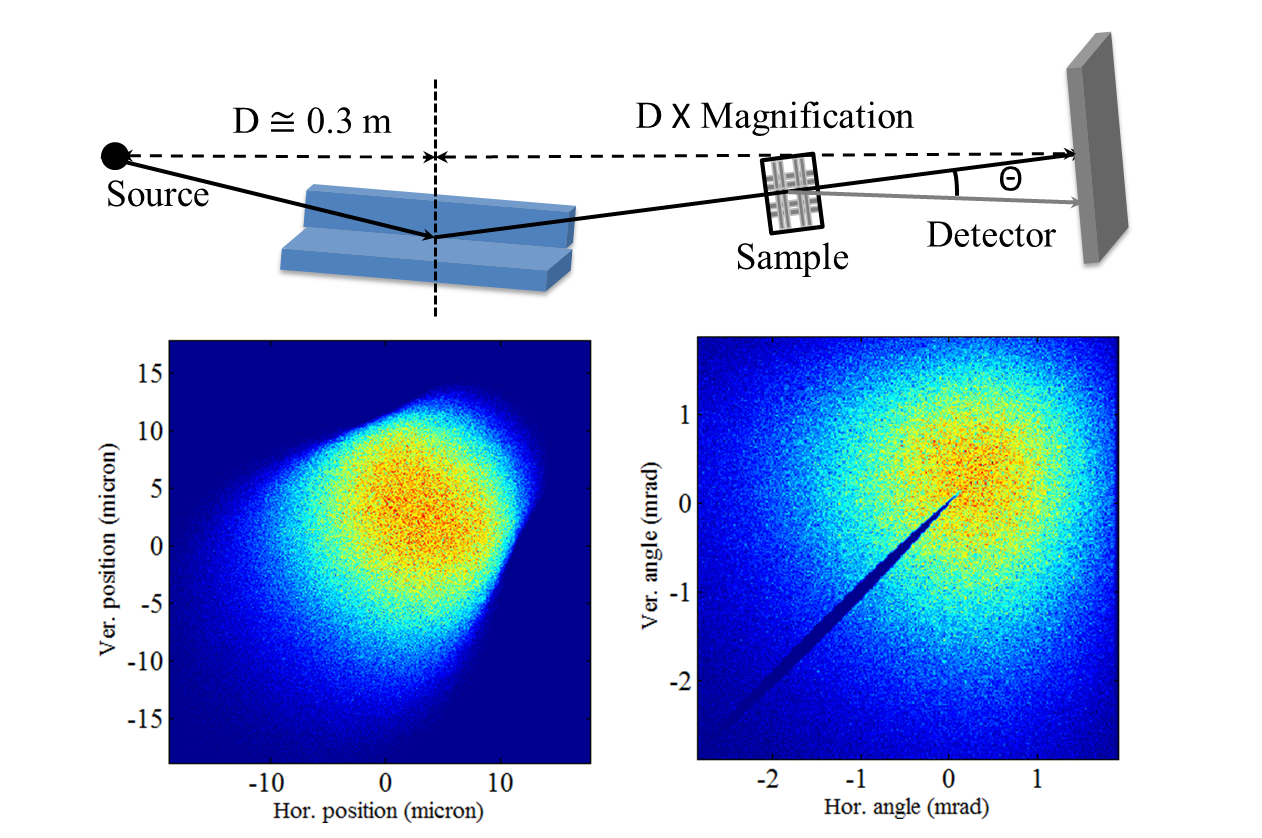}
\caption{Geometry and simulation results for focusing optics used in small angle scattering experiments.  SAXS experiments are well suited to CXLS beam properties, taking advantage of the micron-sized source and higher flux available in a few percent bandwidth. In this arrangment the KB optic produces a factor of 3 magnification to produce a $\sim100\ \um$ converging beam at the sample location on the way to a $10\ \um$ focus at the detector.}
\label{fig:saxs}
\end{figure*}

In order for equation \ref{eq:brilliance} to be valid, both the electron and laser pulse durations should not significantly exceed the laser Rayleigh length $Z_R$.  Additionally, $\sigma_L$ must not be so small that the nonlinear effects begin to degrade the scattered x-ray spectrum.  For $\lambda_L = 515$~nm and $a_{0max} \approx 0.1$, pulse durations on the order of a pico-second and interaction spot sizes of a few microns will be desired to achieve optimum x-ray beam brightness.  Figure \ref{fig:fluxvpulselength} shows numerical results of the effects of laser pulse length and focus size on brilliance and flux.  Ideally, a pulse length less than 1~ps with a near diffraction-limited focus size will produce the highest output.  Practical considerations for laser materials lead us to choose cryo-cooled Yb:YAG which is capable of 2~ps pulse duration with very high average and peak power.  The intended operating points are marked on the plots.  We are performing R\&D on cryo-cooled Yb:YLF that should lead to sub-ps pulses at high average power.  The x-ray pulse time profile is shown in Figure \ref{fig:xrayvstime} with an RMS pulse length of 490~fs, similar to the electron beam.  Because the electrons are propagating at close to the speed of light, the x-ray pulse length depends only on the electron bunch length and for practical purposes is independent of the laser pulse length.

In summary, the optimization of high average flux, high spectral brightness inverse Compton scattering x-ray sources requires electron beams with low emittance ($<200$~nm-rad) and short pulse duration ($<1$~ps), and tightly focused ($<5~\um$), short pulse ($<\textrm{ps}$) lasers.  The x-ray performance resulting from numerical optimization of the ICS source using state-of-the-art laser and accelerator technology is presented in Table \ref{tab:xray}.

The technical aspects of the laser and electron beams and the equipment used to produce them are presented in the sections that follow.

%%%%%%%%%%%%%%%%%%%%%%%%%%%%%%%%%%%%%%%%%%%%%%%%%%%%%%%%%%%%%%%%%%%%%%%%%%

\section{X-ray optics}

A wide array of x-ray beamlines and instruments are enabled by the CXLS, each demanding particular optics whose description is beyond the scope of this article.  However two example optics are presented that match well with the CXLS source properties.  It is important to note that Kirkpatrick-Baez mirrors, particularly the nested version also known as Montel optics, are generally well-suited for both collimating and focusing x-ray beams with the characteristic sizes and divergences produced by this source technology.  The collimating function is illustrated by ray-tracing results shown in Figure \ref{fig:collimation} to prepare the beam for monochromatization.  The source is represented by an axisymmetric Gaussian in angle (4.3~mrad RMS) and a flat-top cylindrical spatial profile ($5.9~\um$ diameter).   The beam which emerges from the optic has collimation of about $10~\mathrm{\mu rad}$ and size of about 1~mm, which are ideal properties for coupling into a standard perfect crystal monochromator with high efficiency.  Such an arrangement would be valuable for most scattering experiments requiring monochromatic beams, such as macromolecular crystallography.

To illustrate a possible application using a focusing geometry, we describe a setup for small-angle x-ray scattering (SAXS), which is a heavily used method to obtain structural information in a variety of fields.  An initial x-ray beamline layout is shown in Figure \ref{fig:saxs}. The diverging x-ray beam is reflected from the mirrors and then transmitted through the sample and focused on the detector.  The mirrors are two perpendicular elliptically bent slabs; the beam must be reflected twice, once from each mirror. After the sample, the scattered beam is detected, while the transmitted unscattered beam is blocked. The full length of the system will be $\sim$120~cm depending on the detector resolution and sample size.

\begin{figure*}[t]
\includegraphics[width=0.90\textwidth]{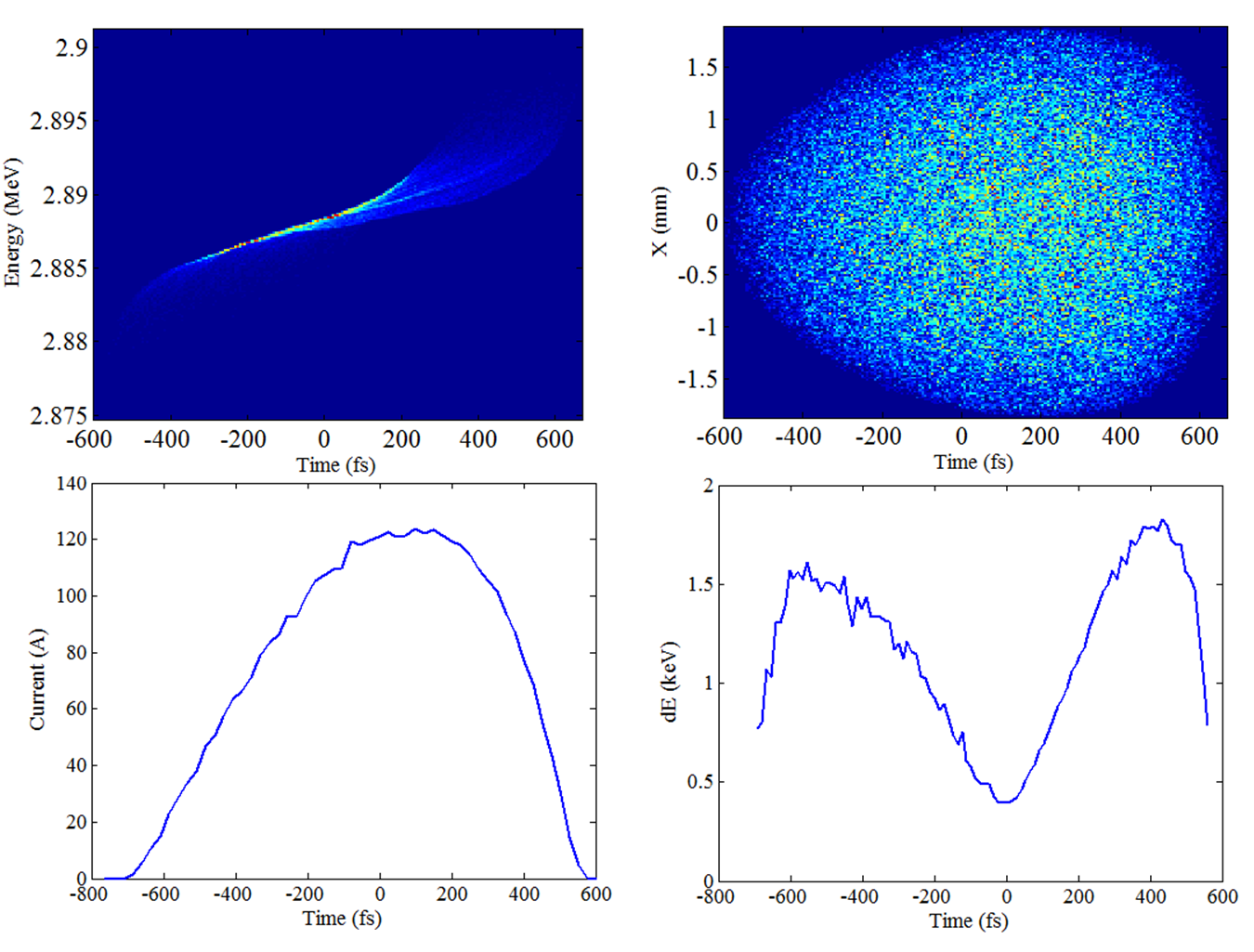}
\caption{Plots of longitudinal electron beam parameters at gun exit vs time relative to bunch center.  Upper left shows time-energy phase space with mean energy of 2.89~MeV and RMS bunch length of 260~fs ($\sim1\degree$ RF).  Upper right shows ellipsoidal distribution resulting from blowout mode dynamics.  Lower left shows 120~A peak current and lower right shows energy spread that is higher in the head tail; however, energy spread remains below 2~keV for all time slices. The bunch length has expanded by a factor of 4 from the cathode laser pulse length as required by blowout mode dynamics.}
\label{fig:eslicesgun01}
\end{figure*}

The intensity of the beam scattered by the sample is measured as a function of the momentum transfer Q, or the scattering angle. At small angles, $Q=4 \pi/\lx \sin \theta/2 \simeq 2 \pi \theta/\lx$. Imagine that the sample is a periodic structure with spacing $d$, resulting in diffraction orders (Bragg peaks) at momentum transfers $Qd=2 \pi n$. The d-spacing can be obtained from distances between the diffraction orders, $\delta Q=2 \pi/d \simeq 0.3~\mathrm{nm^{-1}}$ for spacing = 22~nm, for example. Angular positions of the diffraction orders are determined by $n \lx = 2 d \sin \theta/2 \simeq dθ$. From here, $\Delta \theta = \lx/d \simeq 4.5~\mathrm{mrad}$. In SAXS measurements at APS \cite{settens}, about 20 diffraction orders were measured in such a sample, corresponding to $Q_{max} \simeq 20 \delta Q \approx 6~\mathrm{nm^{-1}}$ and $\theta_{max} \simeq 90~\mathrm{mrad}$. 

For a source-optic distance of 30~cm the beam size at the mirrors is 3 mm assuming beam divergence (FWHM) of 10~mrad. The sample should be placed between the optics and the detector, where the cross-section of the beam is the size of the sample, about $80~\um$. For magnification M=3 the optic-detector distance is 90~cm, thus samples should be placed 2 cm upstream of the detector to be illuminated by the full cross-section of the beam. Using $\Delta \theta \simeq 4.5~\mathrm{mrad}$, the spatial separation between diffraction orders at the detector will be $90~\um$, which can be resolved with a YAG:Ce scintillator and visible light ccd camera.

The resolution of such an instrument in which the beam is focused on the detector is determined by uncertainties of the wavelength and the scattering angle $(\Delta Q / Q)^2 = (\Delta \lx / \lx)^2 + (\Delta \theta / \theta)^2$.  Assuming ICS bandwidth of 5\%,  $(\Delta \lx / \lx)^2 \approx 2.5 \times 10^{-3}$. The geometric contribution is $(\Delta \theta)^2=(b^2+p^2) / L_{SDD}^2 + FE^2$.  Here $L_{SDD}$ is the sample-to-detector distance (2 cm in this example), b is the size of the direct-beam spot on the detector, p is the intrinsic resolution of the detector and $ FE \simeq 5~\mathrm{\mu rad}$ is the figure error of the mirrors. The direct-beam size b is determined by the source diameter and the magnification of the optics. Optimally, the magnification of the optics should by such that $b \approx p$, which determines the magnification M.  Assuming $b \simeq p \simeq 10~\um$, the width of the Bragg peaks $\Delta \theta \approx 0.7~\mathrm{mrad}$. Therefore, the geometric contribution to the resolution $ (\Delta \theta / \theta)^2$ decreases quickly with increasing angles, as  $(\Delta \theta / \theta)^2 = 0.05/n^2$, where n is the diffraction order number. Hence the resolution of the SAXS optical system is mostly determined by the bandwidth of the source.  The full source bandwidth $\Delta Q / Q \approx \Delta \lx / \lx \approx 5\%$ will resolve up to the 20th harmonic, and can be narrowed if higher resolution is desired to reach even higher harmonics.

\begin{figure*}[t]
\includegraphics[width=0.90\textwidth]{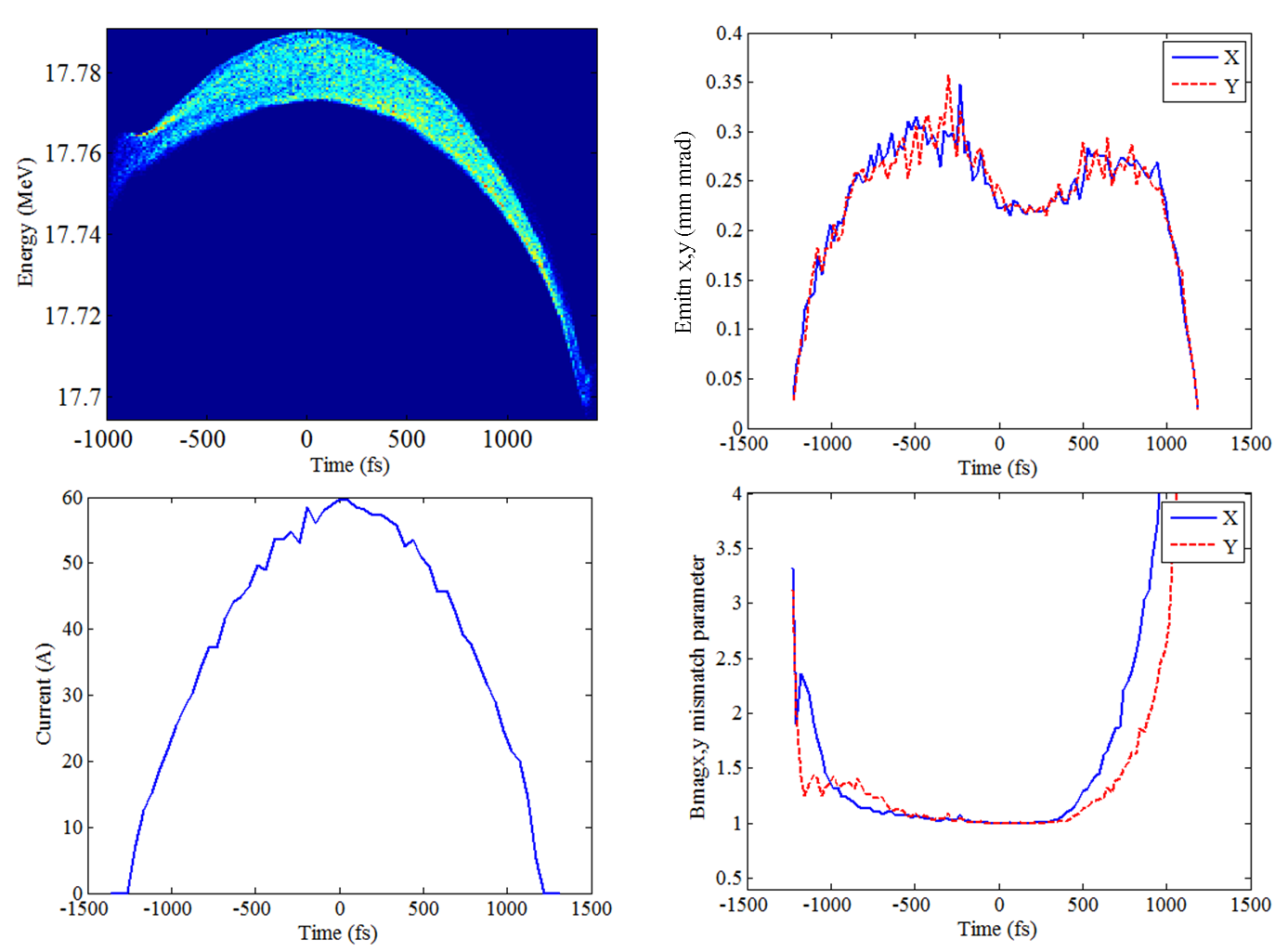}
\caption{Plots of transverse and longitudinal electron beam parameters at IP vs time relative to bunch center.  Upper left shows time-energy phase space with mean energy of 17.9~MeV and RMS bunch length of 580~fs, a factor of 2 longer than at gun exit due to space charge dynamics.  Upper right shows that slice emittance remains low.  Lower left shows 60~A peak current.  Lower right plot shows $\mathrm{B_{mag}}$, which measures how well the different timeslices overlap the central timeslice in transverse phase space.  Ideally $\mathrm{B_{mag}} = 1$ for all slices, but the central high current slices are most important.}
\label{fig:eslicesIP01}
\end{figure*}

We conclude that excellent resolution can be obtained with quite strong scattering signals, given that the incident flux in a 5\% bandwidth is approximately $5 \times 10^{11}$~photons/sec.  One major advantage is the simplicity of the beamline compared to APS, and also the fact that the beamline length is just over a meter.  At APS the sample to detector distance alone is 8~meters. 

%%%%%%%%%%%%%%%%%%%%%%%%%%%%%%%%%%%%%%%%%%%%%%%%%%%%%%%%%%%%%%%%%%%%%%%%%%

\begin{figure*}[t]
\includegraphics[width=0.90\textwidth]{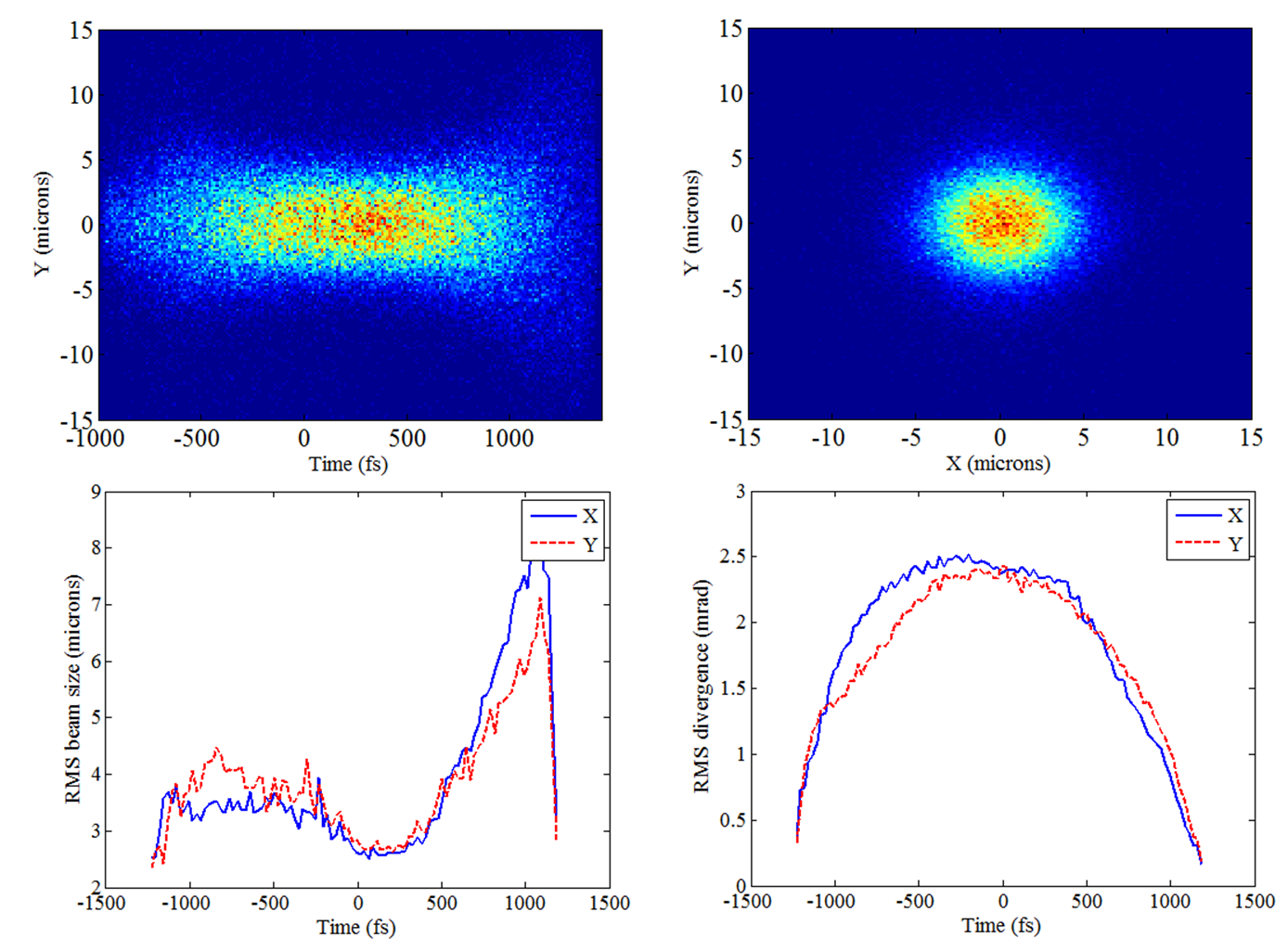}
\caption{Plots of transverse properties of electron beam at interaction point.  Upper left shows density plot of electron position vs time showing most of beam is well focused.  Upper right plot shows time-integrated beam intensity with RMS spot size of $3.5~\um$.  Lower left plot shows variation in focus size of different time slices and lower right plot shows beam divergence vs time.  The electron beam divergence of 2-3 mrad is similar in magnitude the x-ray opening angle. }
\label{fig:eslicesIP02}
\end{figure*}

\section{Electron beam dynamics}
The electron bunch charge should be as large as possible of course to maximize x-ray flux but is constrained by the needs for small emittance and short bunch length in order to produce bright x-rays, as well as charge-dependent effects of beam loading and wakefields in the RF structures. The maximum bunch charge is also limited by the available cathode-laser power and the cathode quantum efficiency.  The emittance requirement $ \exn \approx 0.2~\um$ is set so that the electron divergence at the micron sized focus at the IP has minimal impact on the x-ray bandwidth.  The bunch length during acceleration should be limited to a few RF degrees ($3\degree \approx 1$~ps at 9.3~GHz) so that the energy spread remains small.  The blowout mode of generating an ellipsoidal bunch distribution \cite{luiten} is used for its ability to generate short bunches with a uniform charge distribution that exhibits linear space charge forces thus avoiding emittance growth.  These bunches are well suited for temporal compression and show excellent focusing characteristics for producing micron-sized spots at the interaction point.  The blowout method reduces sensitivity to laser temporal pulse shaping and inhomogeneities in the cathode emission.  Furthermore, because the transverse space charge forces are not only linear, but also identical in each time slice, there is little relative rotation of the timeslices in phase space so that emittance compensation schemes \cite{ferrario} are less critical.

For robustness in the high 140 MV/m RF fields we choose to use a copper cathode with a design quantum efficiency of $5 \times 10^{-5}$ at the 140~MV/m applied RF field.  The available cathode laser power of 20 W is then consistent with producing bursts of 100 bunches in $0.5~\us$ at a repetition rate of 1~kHz.  The bunch charge is also limited by beam loading of the RF fields.  The total charge of 10~nC contained in the bunch train produces a linac beam loading of 16\%.

The initial electron beam emittance is $\exn = \sigma_x \sqrt{(h\nu - \phi_{eff})/3 mc^2}$ where $h\nu = 4.81$~eV is the photon energy of the frequency-quadrupled Yb:KYW cathode laser, $\phi_{eff} = 4.14$~eV is the effective work function~\cite{dowell_2009} of Cu(100) at an applied field of 140 MV/m, and $\sigma_x$ is the RMS beam size.  A laser spot RMS radius of $260~\um$ on the cathode, or an edge radius of $600~\um$ for the parabolic intensity profile, will then produce an initial emittance of $0.17~\um$.  The initial FWHM pulse length is 150~fs set by the UV laser pulse length.

\begin{figure*}[t]
\includegraphics[width=0.90\textwidth]{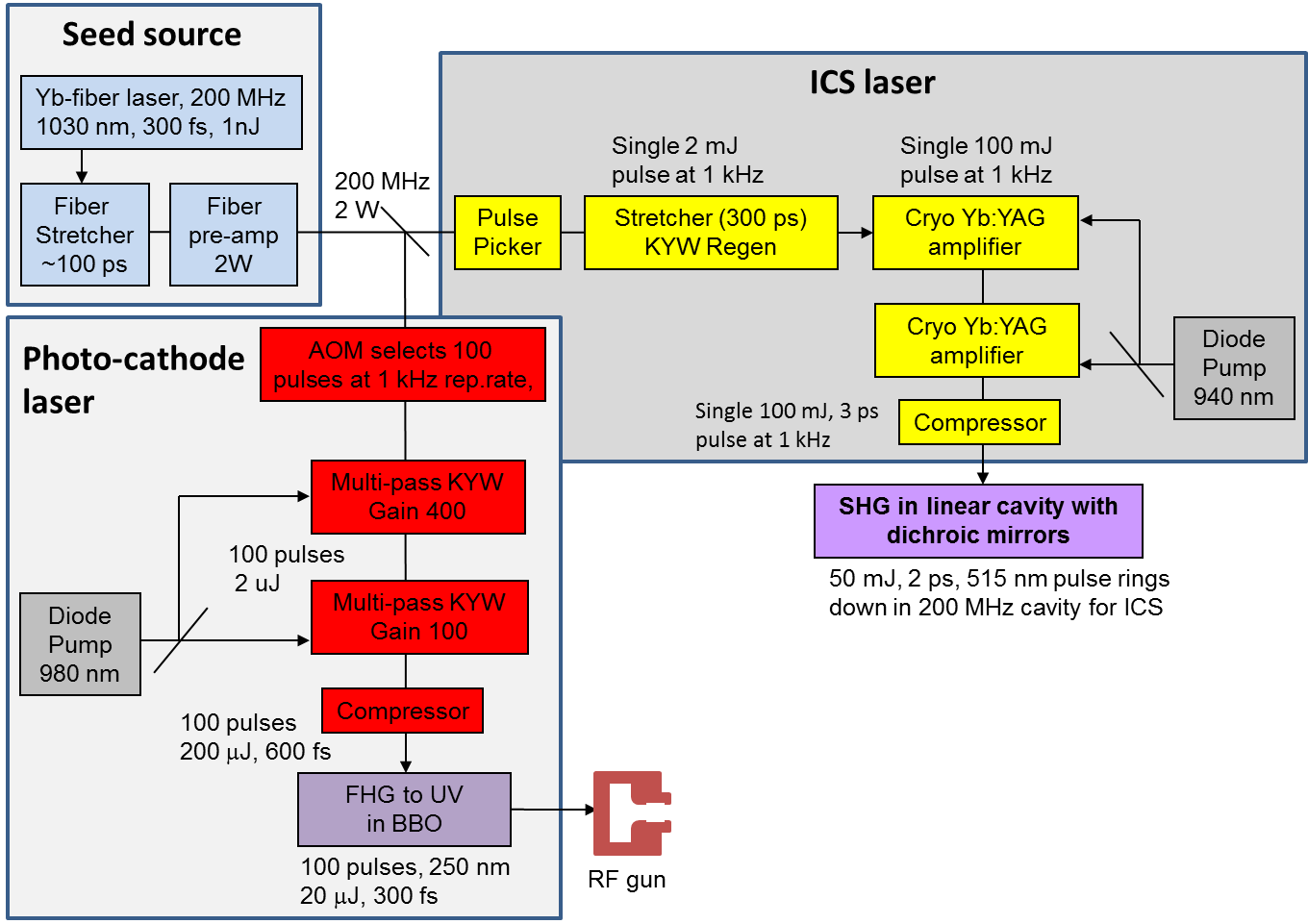}
\caption{Block diagram of the laser systems showing the common oscillator feeding into the amplifiers for the photocathode laser and ICS laser.  The Yb:KYW photocathode laser requires a 100 pulse train at 1~kHz repetition rate for an average output power of 20 W in the IR.  The ICS laser is a cryo-cooled Yb:YAG amplifier that produces a single 100~mJ IR pulse at 1~kHz.  The IR pulse is coupled into a linear ringdown cavity where it is trapped using second harmonic generation and dichroic mirrors \cite{jovanovic}. The 515 nm light then collides with electron bunches that arrive in a train synchronized to the cavity roundtrip time.}
\label{fig:laserblockdiagram}
\end{figure*}

To properly set up the dynamics of the blowout mode the space charge field near the cathode must be much smaller than the applied RF field, but large enough that the bunch length at the gun exit is significantly longer than its initial value. In the thin disk approximation the peak space charge field is $3Q/(2 \epsilon_0 \pi r^2) = 16$~MV/m.  The RF field at the time of emission is $140 \sin(50\degree) = 108$~MV/m satisfying the first condition.  Numerical simulations (Figure \ref{fig:eslicesgun01}) indicate that the second condition is satisfied as the RMS bunch length increases from 65~fs to 260~fs at the gun exit, resulting in a peak current of 120~A.  Figure \ref{fig:eslicesgun01} also shows that the bunch has expanded into the desired ellipsoidal distribution at an energy of 2.9~MeV.  The longitudinal phase space shown in the upper left plot indicates that blowout mode has created a chirped energy distribution necessary for the expansion.  The lower right plot of Figure \ref{fig:eslicesgun01} shows that the slice energy spread varies from a low of 0.5~keV at the beam center to 1.6~keV at the head and tail.  The larger energy spread at the head and tail are consistent with blowout mode dynamics where it is the variation in velocity that creates the ellipsoidal distribution.  Although the peak current at the gun exit is quite high, the bunch will continue to stretch in time up until the laser interaction due to the low energy of the entire machine, resulting in a factor of two lower peak current at the interaction point, see Figure \ref{fig:eslicesIP01}.

\begin{table}[h]
\caption{Cathode laser and initial electron beam parameters. \label{tab:cathode} }
\begin{center}
\begin{tabular}{l c c c}\hline\hline\
Parameter & Value &  Unit \\
\hline
Normalized emittance & $1\times10^{-7}$  & m-rad  \\
Peak current & 120   & Amps  \\
Bunch charge   &  100    & pC  \\
Laser FWHM length & 150 &  fs  \\
Laser temporal shape  &  Arbitrary    &  \\
Laser spatial shape  &  Parabolic    &  \\
Laser edge radius & 0.6  & mm \\
Peak RF field & 140 & MV/m\\
RF phase at emission & 50 & degrees  \\
\hline \hline
\end{tabular}
\end{center}
\vspace{-3mm}
\end{table}

\begin{figure*}[t]
\includegraphics[width=0.80\textwidth]{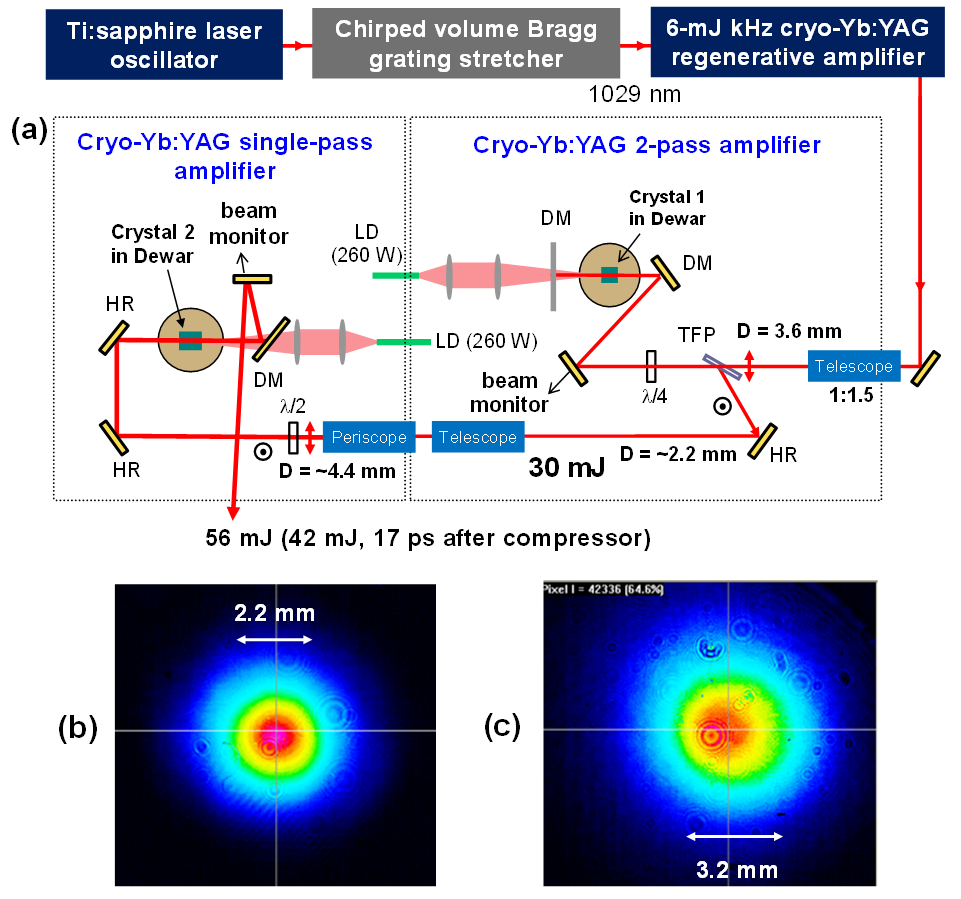}
\caption{Optical layout of the 3-stage kHz picosecond cryogenic Yb:YAG CPA laser (a) and the beam profiles from the 2-pass (b) and the single-pass amplifiers (c) \cite{hong_2014}. The circular fringes in the beam profiles are the measurement artifact from neutral density filters. Crystal 1, 1$\%$ doped, 10-mm-long Yb:YAG; Crystal 2, 2$\%$ doped, 20-mm-long, Yb:YAG; D, diameter; HR, high reflector; DM, dichroic mirror; TFP, thin-film polarizer; $\lambda/2$, half waveplate; $\lambda/4$, quarter waveplate; LD, laser diode at 940~nm.}
\label{fig:rod_YAG_laser}
\end{figure*}

The electron beam exiting the gun is focused by a 6~cm long solenoid with peak field 5~kG to a soft waist of $480~\um$ at the linac entrance to match the Ferrario criterion \cite{ferrario} for emittance correction, i.e. generating a bunch with time slices that are well-aligned in phase space, resulting in a low overall projected emittance at the linac exit.  The short 1~m long standing-wave linac then accelerates the bunch to the energy required for x-ray production, 17.8~MeV in the case of 12.4~keV x-rays. Downstream of the linac, shown in Figure \ref{fig:cadICSassembly}, are a quadrupole pair to match into a 4-magnet chicane that is used primarily to block unwanted stray electrons from entering the laser interaction area using energy and spatial filtering.  The chicane can also be used for bunch compression to produce bunches less than 100 fs in duration, however compression dynamics are not addressed here.  Following the chicane, a short focal length quadrupole triplet focuses the electrons to a small spot at the IP.  The electron beta function $\beta^* = 1.5$~mm at the IP in both transverse dimensions compared with maximum beta functions at the triplet of $\hat{\beta}_x = 73~\mathrm{m}$ and $\hat{\beta}_y = 190~\mathrm{m}$ for a beam size demagnification of a few hundred.  Figures \ref{fig:eslicesIP01} and \ref{fig:eslicesIP02} show the time-dependent variation of electron beam parameters at the IP, including the mismatch parameter $\mathrm{B_{mag}} \ge 1$, which is a measure of how well the phase space of each timeslice overlaps with the phase space of the central slice \cite{loos}.  Ideally, $\mathrm{B_{mag}}=1$ for all slices, but it is most important that the high current slices near the bunch center overlap as in the case shown.  The initial laser and electron beam parameters are given in Table \ref{tab:cathode} and the final electron beam properties are summarized in Table \ref{tab:ebeam}.

\begin{table}[h]
\caption{Electron beam parameters at laser interaction point. \label{tab:ebeam} }
\begin{center}
\begin{tabular}{l c c c}\hline\hline\
Parameter & Value &  Unit \\
\hline
Peak current & 70   & Amps  \\
Normalized emittance & $2\times10^{-7}$  & m-rad  \\
RMS $\mathrm{\delta E/E}$  &  $8\times 10^{-4}$    & --  \\
Energy  &  8-40 & MeV  \\
Bunch charge   &  100    & pC  \\
RMS bunch length & 490 &  fs  \\
Beta function at IP  &  1.5    & mm \\
Beam size at IP & 1.9  & $\um$ \\
Repetition rate & 100 & kHz  \\
Average current & 10 & $\mu$ A \\
\hline \hline
\end{tabular}
\end{center}
\vspace{-3mm}
\end{table}

%%%%%%%%%%%%%%%%%%%%%%%%%%%%%%%%%%%%%%%%%%%%%%%%%%%%%%%%%%%%%%%%%%%%%%%%%%
\section{Laser technologies}

\begin{figure*}[t]
\includegraphics[width=0.90\textwidth]{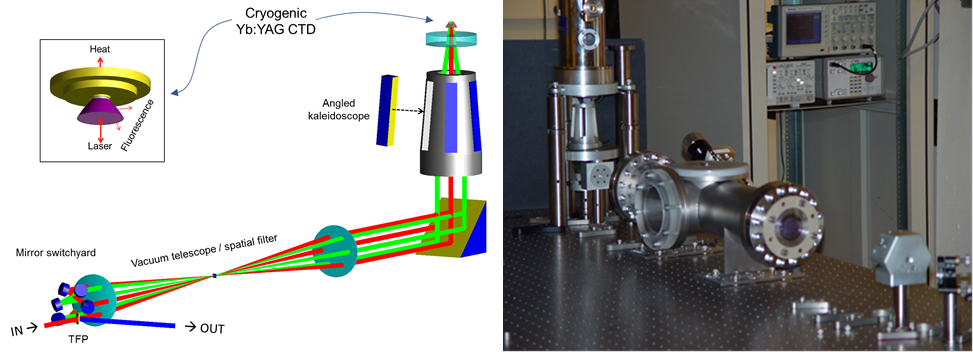}
\caption{The key components in our high average-power high pulse-energy chirped pulse amplifier design are a shaped composite-thin-disk gain-element (inset in the left pane) that is cryogenically cooled and, a passively switched strictly image relayed multipass architecture utilizing a beam-smoothing telescope. A photograph of the hardware is on the right.}
\label{fig:cryo_YAG_laser}
\end{figure*}

An integral part of the ICS source is the laser system which provides both the ICS laser and the photocathode laser. A schematic layout of the laser system,  shown in Figure \ref{fig:laserblockdiagram}, consists of two amplifier chains driven by the same Yb-doped fiber oscillator to achieve optical synchronization. The mode-locked Yb-doped fiber oscillator generates a 1030~nm pulse train with a repetition rate of 200~MHz which sets the frequency for the burst of electron pulses. The spectral bandwidth of the 1~nJ fiber oscillator pulse is $\sim$12~nm, which is broad enough to counteract the effects of gain narrowing in the two amplifier chains and ensure the compressed pulse duration in the sub-ps range. The fiber oscillator pulse is stretched to $\sim$100~ps and pre-amplified to 10~nJ in order to provide enough pulse energy for both main amplifier chains and thereby minimize the gain narrowing. The ICS laser amplifier chain, shown as yellow blocks and described in Section \ref{sec:ICS_laser}, is used to pump an ICS ringdown cavity. It consists of a second stretcher and a commercial Yb:KYW regenerative amplifier (Amplitude Systemes, Inc.) which selects pulses at 1~kHz and amplifies them to 2~mJ followed by cryogenic multi-pass Yb:YAG amplifiers to reach 100~mJ with $\sim$3~ps pulse width after compression. The compressed 100~mJ pulse is coupled into the ringdown cavity, described in Section \ref{sec:ringdown_cavity}, by passing through a dichroic mirror and frequency doubling. This cavity allows for 100 interactions with a single laser pulse greatly increasing the ICS x-ray flux. The photo-cathode laser amplifier chain, shown as red blocks and described in Section \ref{sec:Photocathode_laser}, is used for the photoinjector of the RF gun. This laser requires a burst mode format of 100 pulses at 1~kHz. Two multipass Yb:KYW amplifiers increase the IR pulse energy to 200~$\mu$J for an average power of 20~W. The output is frequency quadrupled to produce the UV pulse for the photo-injector.

\subsection{ICS collision laser}
\label{sec:ICS_laser}

The design of the ICS collision laser is based on high-energy picosecond cryogenic Yb:YAG laser technology recently developed at MIT. The energy scaling of picosecond laser pulses at high repetition rates in the kHz range is a nontrivial task because pulse stretching is not as easy as with femtosecond laser pulses and thus the amplified pulse fluence can easily reach the damage threshold of optical materials and coatings. Recently, new high-power ultrafast laser technologies based on Yb:YAG gain media at 1030~nm have been explored at MIT \cite{Fan, Hong_OL2} and other research groups \cite{Akahane, Metzger}. Here, two approaches are considered for the ICS laser amplifier design: 1) a conventional rod-type cryogenic Yb:YAG amplifier \cite{Hong_OL2} and 2) a composite-thin-disk cryogenic Yb:YAG amplifier with strict image relay \cite{zapata_2013}. 

First, we present a high-energy rod-type cryogenic Yb:YAG amplifier operating at kHz repetition rate \cite{hong_2014}. The layout of a multi-stage cryogenic Yb:YAG chirped-pulse amplification (CPA) laser system is illustrated in Figure \ref{fig:rod_YAG_laser}(a). The output from a Ti:sapphire oscillator is pre-amplified in Yb-doped fiber amplifiers, stretched by a chirped volume Bragg grating (CVBG) pair to $\sim$560~ps with 0.7~nm of bandwidth at 1029~nm, and then amplified by a kHz 6-mJ regenerative amplifier \cite{Hong_OL2} and two multipass amplifiers that are all based on cryogenically cooled Yb:YAG. We used a 1$\%$-doped 10-mm-long Yb:YAG crystal in the first 2-pass amplifier and a 2$\%$-doped 20-mm-long crystal in the second single-pass amplifier. Each crystal has 2~mm of an undoped YAG end cap on the pumping side. The crystals are indium bonded to a heat sink and cooled to 77~K by liquid nitrogen in a vacuum chamber connected to an auto-refilling system. The beam size and the divergence at each stage have been carefully matched to the pump beams using telescopes. The periscope after the 2-pass amplifier helps compensate the thermally induced astigmatism induced by both amplifiers. The maximum energy from the second and third amplifiers is 30 and 56~mJ, respectively, with excellent beam profiles, as shown in Figures \ref{fig:rod_YAG_laser}(b) and (c). The pump power from two fiber-coupled continuous wave 940~nm laser diodes at the maximum output energy are both 240~W, while the absorption is $\sim$75$\%$ at Crystal 1 and $\sim$95$\%$ at Crystal 2, respectively. The amplified pulse with a spectral bandwidth of 0.2~nm is compressed to 17~ps using a multi-layer dielectric grating pair with a throughput efficiency of 75$\%$, delivering a compressed energy of 42~mJ. This is one of the most powerful high energy picosecond lasers operating at kHz repetition rates. 

\begin{figure*}[t]
\includegraphics[width=0.90\textwidth]{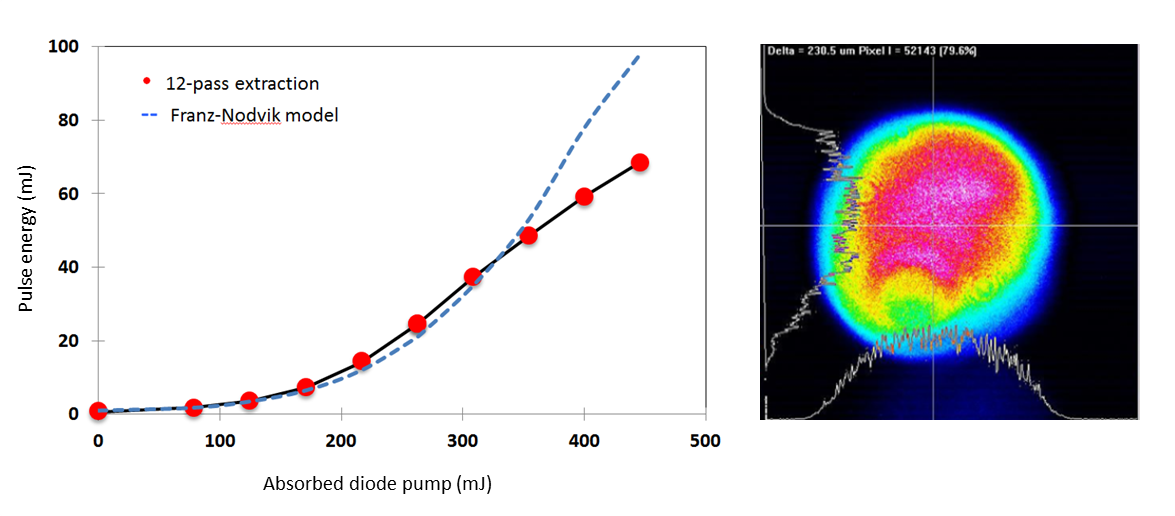}
\caption{(left) Achieved pulse energy at 200 Hz using a high-energy high-power cryogenic composite-thin-disk Yb:YAG amplifier and (right) the near field beam pattern at 60 mJ pulse energy.}
\label{fig:cryo_YAG_experiment}
\end{figure*}

It should be noted that the pulse duration before  compression is $\sim$150~ps which is shortened from ~560~ps due to the gain narrowing in the regenerative amplifier. Currently, the amplified pulse energy of 56~mJ is limited by nonlinear phase accumulation (B-integral). Our simulations \cite{Fu_JOSAB} show that 100~mJ of energy can be obtained if the stretched pulse is maintained to be $\sim$300~ps using a room-temperature Yb:YAG or Yb:KYW regenerative amplifier in which the gain narrowing effect is significantly reduced due to $\sim$10 times broader emission bandwidth than that at cryogenic temperature.  In this case, the compressed pulse duration can be reduced to ~3 ps as well due to a broader spectral bandwidth.

It is desirable to have a more compact Yb:YAG amplifier that also enables higher energies in the future.  To meet those goals we have demonstrated a high-energy high-power cryogenic composite-thin-disk Yb:YAG laser with a strictly image relayed multipass architecture \cite{zapata_2013}. Our goal is a 100~W average power laser system with 1~kHz, 100~mJ pulses of $\sim$3~ps duration to drive the ICS laser ringdown cavity. The heart of our laser driver is a diode pumped cryogenic composite-thin-disk that can operate at higher gain-per-pass than a traditional room temperature thin-disk (inset in Figure \ref{fig:cryo_YAG_laser}). On the cooled face, the laser-grade high-reflector is in intimate contact with a cryogenically cooled heat-spreader through soldering,. The opposite face of the thermally-loaded gain-sheet is diffusion bonded to an index matched cap of undoped YAG. The function of the undoped cap is to dilute fluorescence diminishing the influence of amplified spontaneous emission (ASE) and dramatically enhancing energy storage of inverted Yb$^{3+}$ ions. The edges are fashioned to eject fluorescence; furthermore, the much stiffer gain element affords resilience to thermo-mechanical deformations for excellent beam quality.

We utilize a chirped pulse amplification technique \cite{strickland_1985}, in which the seed pulses for the amplifier chain are generated from a femtosecond Yb-doped fiber laser \cite{hong_2008} and stretched with conventional gold-ruled (Horiba) gratings. The seed is boosted to $\sim$5~nJ with an in-line Yb-doped fiber amplifier before injecting a commercial Yb:KYW regenerative amplifier (Amplitude Systemes) that outputs 2~mJ, $\sim$0.3~ns pulses at repetition rates up to 1 kHz. The output pulses from the regenerative amplifier are then injected into the cryogenically cooled composite-disk multi-pass amplifier utilizing  the greatly improved thermo-mechanical, thermo-optical and spectroscopic advantages of operating Yb:YAG at cryogenic temperature \cite{hong_2008,ripen_2005}. To increase bandwidth, the multi-pass amplifier operates at 130~K bringing the pulse energy to 100~mJ before being compressed with dielectric-coated gratings (Plymouth Gratings). The demonstrated performance for the cryogenic Yb:YAG laser is shown in Figure \ref{fig:cryo_YAG_experiment}.  Increasing the stretching factor will allow for the amplification of higher energy pulses.

\begin{figure}[t]
\includegraphics[trim=0.1cm 0.1cm 0.1cm 0.1cm, clip=true, width=0.45\textwidth]{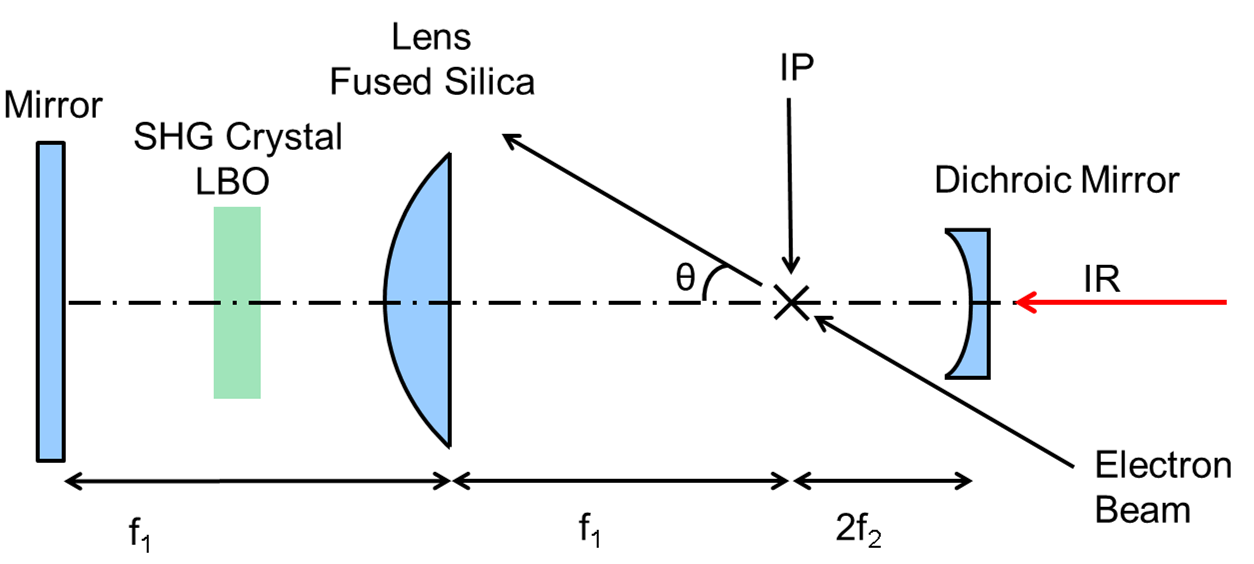}
\caption{Schematic of the linear ringdown cavity with the interaction point labeled IP corresponding to both the electron beam and laser beam waist. The electron beam is at an angle $\theta=50$~mrad w.r.t. the laser beam. }
\label{fig:linearcavityschematic}
\end{figure}

\begin{figure}[t]
\includegraphics[width=0.45\textwidth]{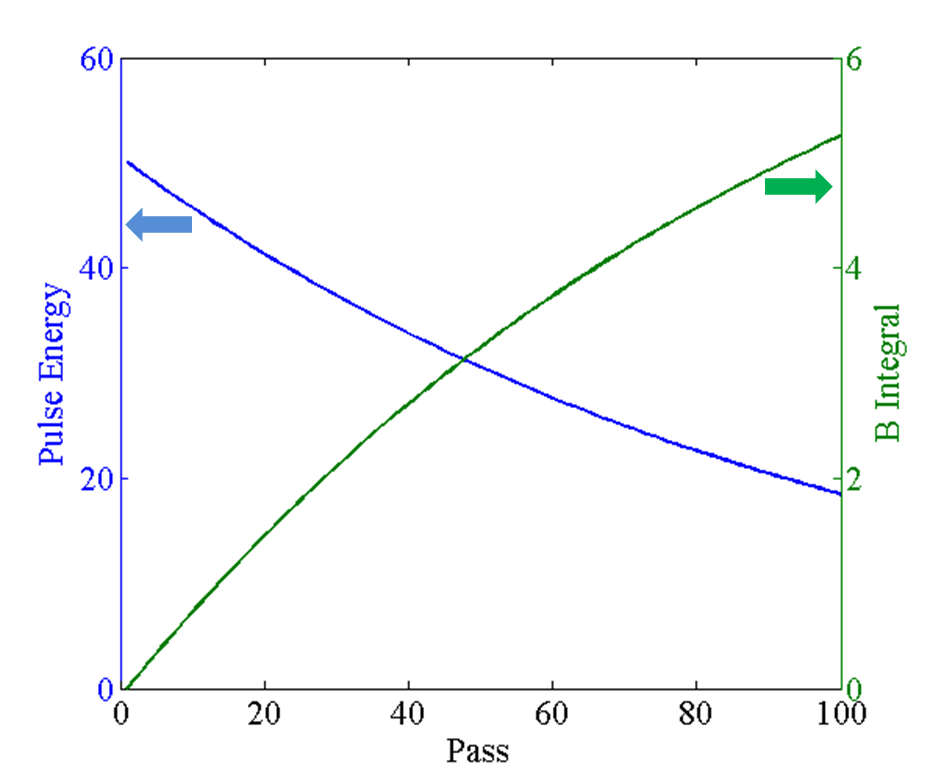}
\caption{Pulse energy and B-integral as a function of pass number in the ringdown cavity.}
\label{fig:ringdown}
\end{figure}

\subsection{Ringdown cavity}
\label{sec:ringdown_cavity}

\begin{figure*}[t]
\includegraphics[width=0.90\textwidth]{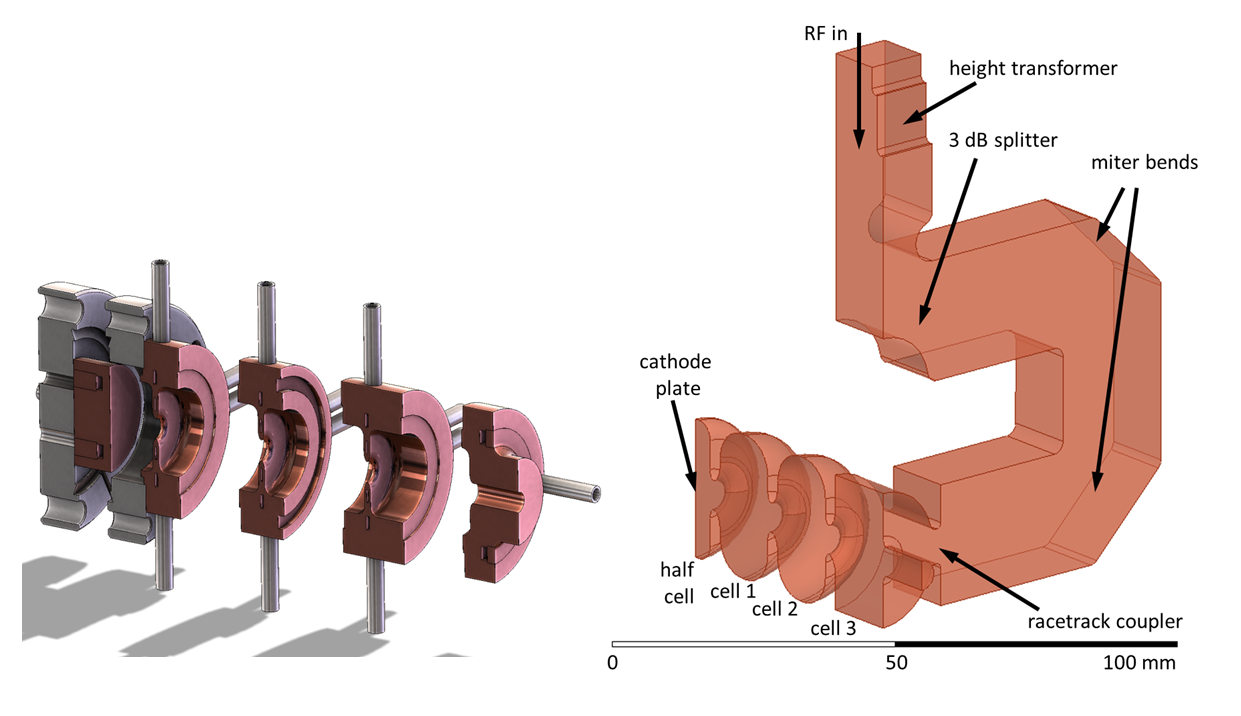}
\caption{Left side shows assembly of photoinjector cells including extensive cooling channels needed for 3 kW heat load.  Right side shows CAD model of the 3.5 cell cavity with coupling and waveguide.  The final cell acts as a coupling cell.}
\label{fig:Valerygun}
\end{figure*}

\begin{table}[hb]
\caption{ICS linear ringdown cavity. \label{tab:lasercavityparameters} }
\begin{center}
\begin{tabular}{l c c c }\hline\hline\
Parameter &  1030 nm  & 515 nm & Unit \\
\hline
Repetition Rate & 200 & 200 & MHz  \\
Focal Length (f1) & 34.5 & 34.5 & cm  \\
Focal Length (f2) & 3 & 3 & cm  \\
$w_0$ & 4.1 & 3 & $\mu$m \\
$w_\text{lens}$ & 26.66 & 18.85 & mm\\
$w_\text{dichroic mirror}$ & 4.64 & 3.28 & mm\\
Pulse Energy  & 100 & 50 & mJ \\
Pulse Width & 2.8 & 2 & ps \\
Peak Surface Intensity & 26.15 & 73.97 & GW/cm$^2$ \\
Peak Energy Density & 73.2 & 148 & mJ/cm$^2$ \\
Lens Thickness & 2.8 & 2.8  & mm  \\
SHG Crystal Thickness & 1.57 & 1.57  & mm  \\
B-Integral  & -- & 5.29 & -- \\
Passes  & 1 & 100 & -- \\
\hline \hline
\end{tabular}
\end{center}
\vspace{-3mm}
\end{table}

%\begin{figure}[t]
%\includegraphics[width=0.45\textwidth]{injector_assembly02.png}
%\caption{Assembly of the photoinjector cells showing cooling channels.}
%\label{fig:injectorassembly}
%\end{figure}

A linear ringdown cavity was selected because of its ability to produce a small micron-sized and symmetric focal spot at the interaction point. Second harmonic coupling into the cavity \cite{jovanovic} is a good solution for high power pulses.  The linear cavity, shown schematically in Figure \ref{fig:linearcavityschematic}, is arranged for conjugate image relay. The round trip length of the cavity is 1.5~m which corresponds to the repetition rate of the electron bunches in burst mode.  The cavity loss is on the order of 1\% with HR/AR coatings contributing less than 0.2\% loss per element \cite{Newport_HR,qtf_AR} and the LBO crystal contributing 0.2-0.5\% loss \cite{castech_lbo}. This will allow for efficient interaction with all 100 of the electron bunches produced in burst mode. The pulse energy and B-integral per pass are shown in Figure \ref{fig:ringdown}. In order to remain synchronized with the electron bunch burst, the cavity roundtrip frequency will be locked to the fiber oscillator which seeds the ICS collision laser and the photocathode drive laser. The accumulated temporal offset should be held below 0.5~ps given the laser pulse and electron bunch width, which corresponds to a cavity stability of 1.5~$\mu$m. The 100~mJ IR pulse is coupled into the cavity through a dichroic mirror. After coupling into the cavity the IR pulse is up-converted to 515~nm via second harmonic generation in LBO \cite{hong_2009}. The target SHG conversion efficiency is 50\% \cite{zhang_1998, hong_2009} occurring during two passes of the IR pulse through the LBO, in order to minimize the losses and undesired non-linearities during the ring down of the cavity. The residual IR pulse is removed through the dichroic mirror from which it was coupled into the cavity. A Faraday isolator is used to separate the input and residual IR beam to prevent potential laser damage.

Due to the large number of round trips and the significant pulse energy, pulse filamentation due to small-scale self-focusing is a significant concern for an optical cavity which contains a fused silica lens and an SHG crystal. The susceptibility of the cavity to filamentation is determined by the accumulated B-integral \cite{hunt_1989,hunt_1993}, shown in Figure \ref{fig:ringdown}, which should ideally be kept below 1 radian. Due to the large B-integral, spatial filtering in the form of an iris at the interaction point will be used to remove the higher-order content which is produced by self focusing. Alternate materials such as BBO for the SHG crystal and CaF2 for the lens are being considered to further reduce the B-integral. Detailed laser cavity parameters are shown in Table \ref{tab:lasercavityparameters}. 

\begin{figure*}[t]
\includegraphics[width=0.70\textwidth]{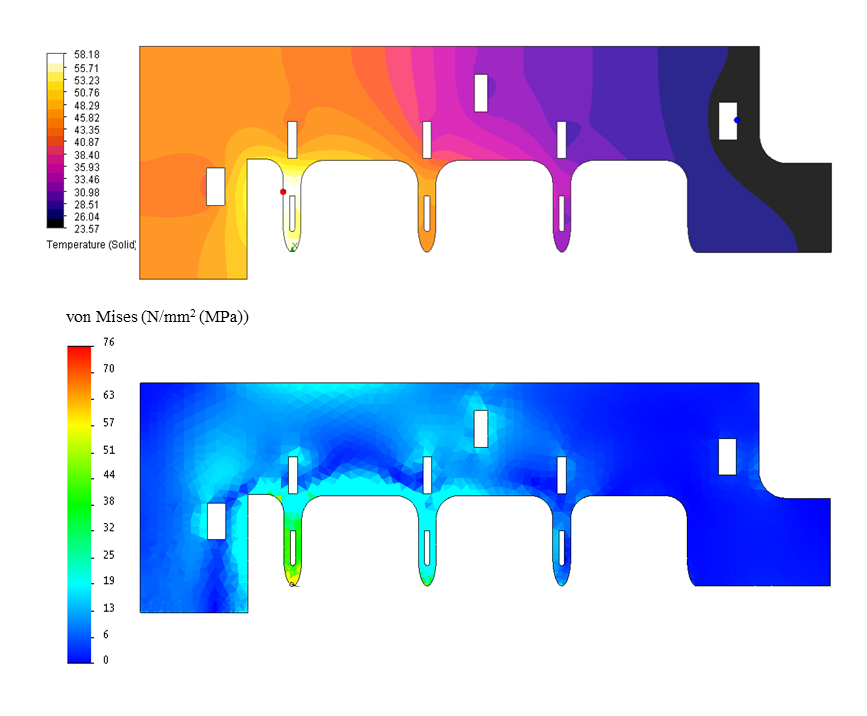}
\caption{Finite element models of the RF photoinjector used for thermal and mechanical analysis.  The upper plot shows temperature at a 3 kW average field corresponding to 3~MW peak and $1~\us$ pulses at 1~kHz repetition rate for a 0.1\% duty cycle.  The maximum temperature rise is $35$~$\degree$C.  The lower plot shows the induced thermal stresses.  There is one small area of maximum stress greater than 50~MPa at the tip of the first iris.  Further optimization, in the form of iris modifications, will reduce this peak stress below that limit.}
\label{fig:gun_heatload}
\end{figure*}

\subsection{Photocathode drive laser}
\label{sec:Photocathode_laser}

Two significant challenges need to be addressed by the photocathode laser. First, the burst-mode pulse format of 100 electron bunches at 1~kHz and 100~pC requires a significant average power of 20~W. Second, in order for the RF photo-injector to operate in the blowout regime the UV laser pulse width needs to be close to $\sim$100~fs.
 
The photo-cathode laser is based on Yb:KYW multipass amplifiers. The pulse bursts are selected by an AOM pulse picker, as illustrated in Figure \ref{fig:laserblockdiagram}. The Yb:KYW gain medium has a higher gain cross-section than the Yb:YAG crystal at room temperature, while having a broad emission bandwidth ($\sim$16~nm) to support sub-ps pulse amplification. It is suitable for producing moderately high average power and low energy per pulsse ($\sim$mJ level) without cryogenic cooling technology. The regenerative amplifier described in the ICS laser section provides a very high gain in a single stage but it cannot amplify the pulse train as needed for a photo-injector. Therefore a multi-stage multipass amplifier with a proper gain control is a straightforward way of obtaining high-power burst of optical pulses. In our design, we set the gain in the first multipass Yb:KYW amplifier to $\sim$400 to obtain 2~$\mu$J of energy from individual pulses and that in the second amplifier to $\sim$100 to obtain 200~$\mu$J of energy. Each burst contains 100 pulses totaling 20~mJ of energy, reaching 20~W of average power at 1~kHz repetition rate. The control of gain narrowing is important to maintain the final spectral bandwidth broader than 3~nm to compress the pulses to $\sim$500~fs. Our calculation shows that the seed spectral bandwidth of $\sim$12~nm with a gain of 4$\times$10$^4$ will result in in a bandwidth of 4.5~nm for the amplified pulses, which supports pulse compression to $\sim$300~fs. The compressed pulses can be frequency quadrupled into UV pulses in two BBO crystals via cascaded SHG with a conversion efficiency of 10$\%$. Finally, we can obtain a kHz UV burst containing 150~fs pulses with 20~$\mu$J of energy, which is an ideal photo-cathode source for an RF gun.

%%%%%%%%%%%%%%%%%%%%%%%%%%%%%%%%%%%%%%%%%%%%%%%%%%%%%%%%%%%%%%%%%%%%%%%%%%

\section{Accelerator RF structures}

We have chosen to work at 9.3~GHz RF frequency (X-band) due to the compact size it enables and the recent availability of MW power klystrons at that frequency paired with compact solid-state modulators.  Lower frequency structures such as S-band or C-band were ruled out on the basis of their size, cost and power requirements.  Although SLAC has developed structures at the nearby frequency of 11.424~GHz, and an ICS source of gamma rays for nuclear studies has been designed at LLNL using that technology~\cite{gibson_2010,albert_2011b}, the only power sources available are large klystrons meant for higher energy accelerators.  The many lessons learned from developing high gradient X-band structures and the LCLS photoinjector at SLAC have been applied to the photoinjector and linac cavity designs.  For the injector these considerations include adequate frequency separation of nearby modes,  compensation of the dipole and quadrupole fields in the coupling cell, and low pulsed heating temperature rise in the structure.  A key design goal is to reach the highest possible efficiency in the structure design in order to enable operation of all accelerator structures with a single 6~MW klystron, and also enabling kHz repetition rates to get to the desired high average x-ray flux.  The subsections below describe the RF gun and linac.  A third RF structure will be used for characterizing the electron beam longitudinal phase space.  It is a standing-wave 5-cell TM110 deflecting mode cavity at 9.3 GHz requiring a few hundred kW of RF power~\cite{nanni_2014}.

\subsection{RF photoinjector}
The photoinjector produces the electron beam and provides initial acceleration to relativistic energy.  Its critical job is to accelerate a short bunch of electrons from rest at the cathode to a few MeV while maintaining the small beam emittance, low energy spread, and a short bunch length.  To accomplish this it requires high RF fields of 100~MV/m or more with low focusing aberrations and high stability.  Thermal loading of the copper structure sets the maximum field strength in the high repetition rate regime.  We have investigated x-band structures having 1.5 cells, 2.5 cells, and 3.5 cells for their ability to produce a beam of several MeV with high cathode gradient, moderate thermal loading, and low RF power demand.  The 3.5 cell gun shown in Figure \ref{fig:Valerygun} outperforms the shorter structures in thermal loading while maintaining a high gradient.  The higher exit energy is also beneficial for reducing space charge effects and producing a less divergent beam.  This structure does require higher power to maintain the high gradient of 140~MV/m on the cathode, needing 3~MW peak, but that is an acceptable power budget for the single klystron.

\begin{figure*}[t]
\includegraphics[width=0.8\textwidth]{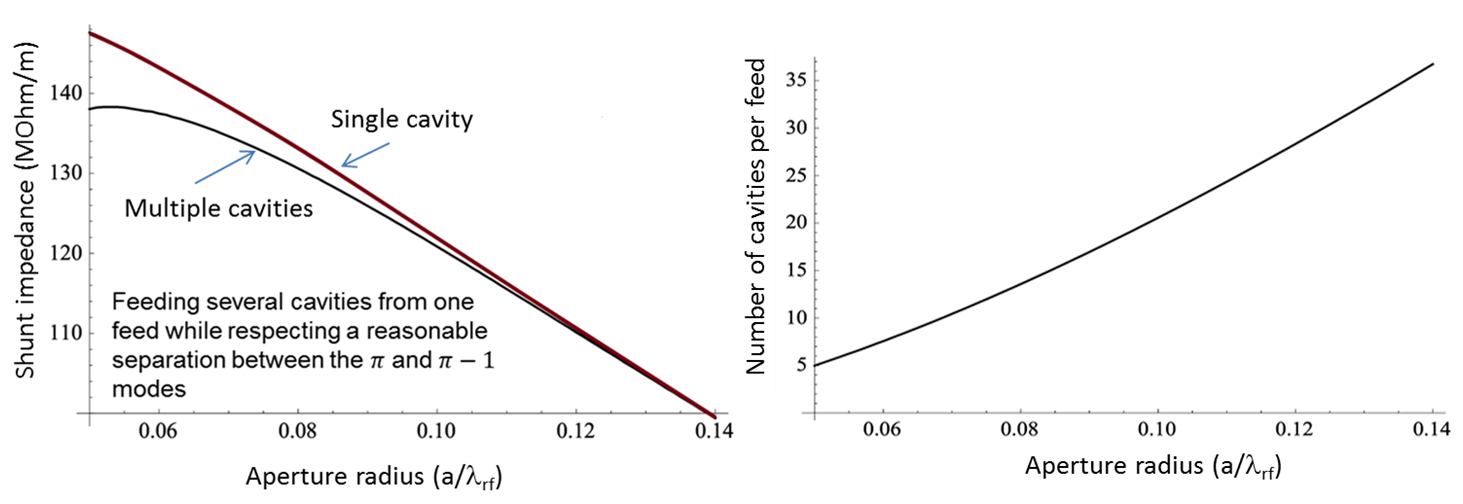}
\caption{Aperture constraints for feeding more than one cavity at a time. (Left) Comparison between shunt impedance for an optimization done on a single cavity and an optimization done on a set of cavities with adequate spacing between the $\pi$-mode and the nearest neighbor mode. (Right) The number of cavities allowed for the given coupling due to the aperture after the optimization.}
\label{fig:shuntimp}
\end{figure*}

The cell layout differs in several respects from other RF photoinjectors.  The ``half'' cell containing the cathode is significantly shorter than $\lambda_{RF}/4$ with the iris center just 5 mm from the cathode.  This is to reduce the electron transit time allowing cathode emission at phases closer to the peak RF field.  The phase of emission depends on the charge desired (in order to cancel the space-charge energy chirp) and is 50 degrees (peak field = 90 degrees) for 100 pC.  This compares with e.g. the LCLS injector phase of 30 degrees which would have a field 50\% lower at emission for the same peak value.  At higher RF frequencies the transit time effect is even more pronounced resulting in the shorter half cell.  The electron bunch is short enough (sub-ps) that phase-dependent RF focusing \cite{kim_rfgun} is not a significant source of projected emittance growth.  We designed the field balance among the cells to give the highest fields (140 MV/m peak) at the cathode and in cell 1 with the peak field falling to 90\% of that in cell 2 and to 45\% in cell 3, which acts as the coupling cell.  The higher fields are important in the early cells to accelerate to relativistic energy to overcome space charge effects.  They are less critical in the last cells.  As seen in Figure~\ref{fig:Valerygun} cell 3 is a coupling cell with a race track shape \cite{Haimson_1999,Nantista_2004} to cancel the quadrupole moment of the RF fields and dual waveguide feeds to cancel the dipole moment.  The lower field in this coupling cell also helps avoid distorting the beam distribution and lowers pulsed heating on the coupling slots.  The four cells can support four different modes.  SUPERFISH studies show the nearest mode is 18~MHz from the desired pi mode, well outside the 2.2~MHz resonance width given by the cavity loaded quality factor of 4200, assuming critical coupling.

\begin{figure*}[t]
\includegraphics[width=0.80\textwidth]{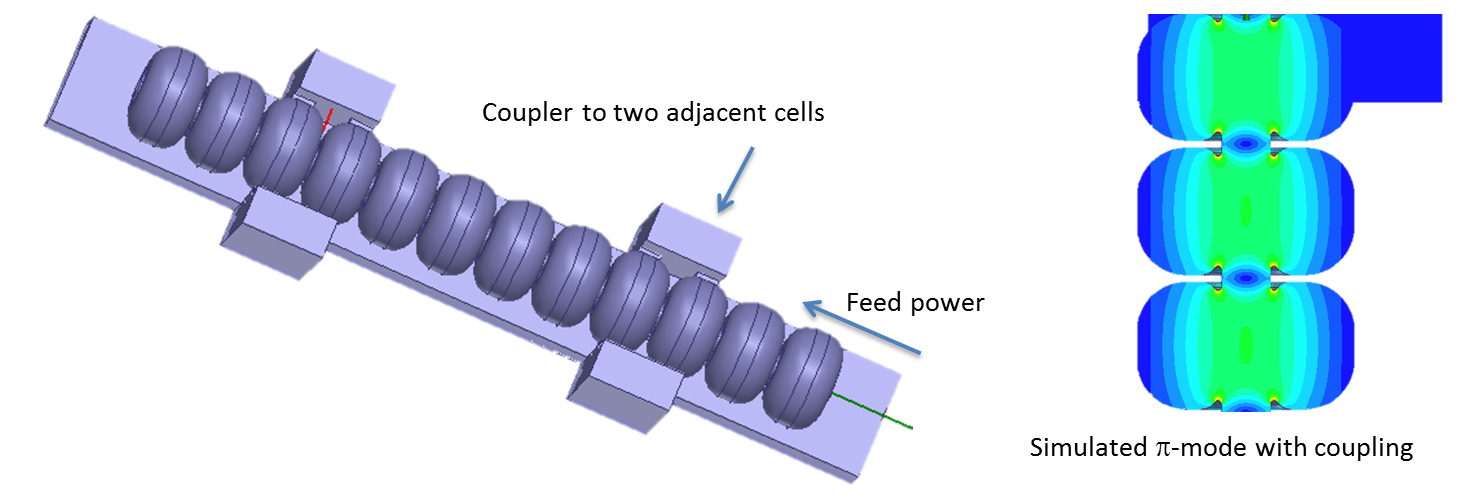}
\caption{High efficiency standing-wave linac structure with distributed coupling.  Two sections each 10 cm long are shown.  Structures would be stacked into a 1.3 m long linac providing approximately 21 MeV energy gain for 1.6 MW input power.   Right side shows $\pi$-mode with coupling.}
\label{fig:linac01}
\end{figure*}

Operating at 1~kHz with $1~\us$ long pulses ($0.5~\us$ flat top) results in a thermal load of 3~kW.  Figure~\ref{fig:gun_heatload} shows the resulting temperature rise and thermal stresses.  The water cooling channels are also visible.  Due to the large thermal load in the cavity, targeted water channel placement is required to minimize temperature rise and thermal stress. Water channels with a total flow rate of 0.21~lps will operate with a pressure drop of 0.45~atm and a temperature rise of 3-4~$^o$C between the inlet and outlet. The RF cavity temperature will be regulated within $\pm$0.1~$^o$C in order to keep the gun cavity on resonance. Feedback on RF phase from a pickup measuring the reflected power from the gun will stabilize the cavity phase. At peak thermal load the inlet temperature and velocity will be 15~$^o$C and 4~m/s, respectively. The assembly will consist of five separate sub-assemblies, brazed or diffusion bonded together. Each section has integral water cooling channels with no joints between water and vacuum.

\begin{table}[h]
\caption{Parameters for linac and photoinjector structures. \label{tab:rfstructures} }
\begin{center}
\begin{tabular}{l c c c}\hline\hline\
Parameter & Photoinjector & Linac & Unit \\
\hline
Length & 5.3 & 1280   & cm  \\
Number cells & 3.5 & 80  & --\\
Shunt impedance & 130  & 123 & M$\Omega$/m  \\
R/Q  &  4.8 & 11.3 & kOhm/m  \\
Q unloaded  & 8,400  & 11,000 & --  \\
Energy gain  &  3.2 & 21 & MeV  \\
RF Power   &  2.9 & 1.6    &  MW  \\
Peak wall E-fld & 170 & 52 &  MV/m  \\
Peak wall H-fld  &  266 & 51   & kA/m \\
Iris diameter  &  6 & 6.4   & mm \\
\hline \hline
\end{tabular}
\end{center}
\vspace{-3mm}
\end{table}

\begin{figure*}[t]
\includegraphics[width=0.75\textwidth]{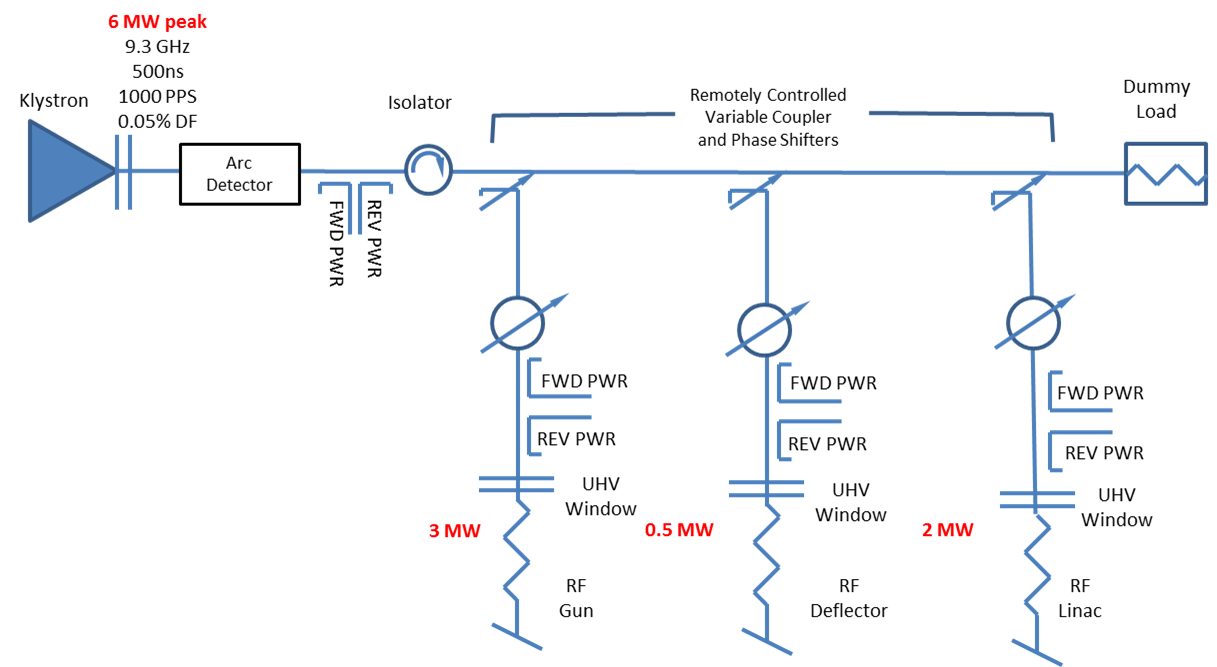}
\caption{Layout of RF waveguide equipment.  Power is divided from the single klystron into 3 arms for the photoinjector, linac, and deflector cavity.  The phase and amplitude in each arm is individually controlled allowing flexibility in the power distribution.}
\label{fig:rfwaveguide}
\end{figure*}

\subsection{RF linac}

The linac design \cite{tantawi_TBB} is based on an invention \cite{tantawi_patent2014}  that distributes the coupling over the cells of a standing-wave accelerator structure \cite{tantawi_2006}. The idea of feeding each cavity has existed for some time \cite{schaffer_1965,sundelin_1977,yu_2010}; however, until now there has been no practical electron linac implementation that allows such a topology to exist. What this invention provides is a practical implementation of a microwave circuit that is capable of feeding individual or multiple cavities separately. This circuit is designed in such a way that the coupling between cavities is minimized. Hence, individual cavities can be optimized without the constraint usually applied from the coupling between adjacent cavities. This has the benefit of more efficient designs that consume less rf power. Therefore, the overall cost of building a linear accelerator system for any application is substantially reduced.  
Indeed, benefiting from the advantages of distributed feeding does not require feeding every single cavity in the structure. A significant advantage can be retained while increasing the number of cells per feed arm from every individual cell to multiple cells. Despite an increase in the number of cells fed per feed arm the shunt impedance would be very high. This is demonstrated in Figure \ref{fig:shuntimp} which shows an analysis of the shunt impedance versus number of cells being fed. The analysis shown assumes the number of cells that are being fed by a single feed arm is limited by the axial modal density. Hence, the number is determined by the proximity of the $\pi$-mode to the nearest neighbor mode, and the need for this separation to be more than the band width determined by the quality factor of the $\pi$-mode. We choose to feed 6-cell segments with 2 independent RF couplers so that 3 cells are coupled.  The resonant modes are at 9.273, 9.291, and 9.300 GHz. The 9 MHz mode separation is well outside the 1.7~MHz resonance width given by the loaded quality factor of 5500 assuming critical coupling with intrinsic quality factor of 11,000.  The analysis is done for cavities operating at 9.3~GHz, but it can be done at any other frequency as well to yield similar results. 

The distribution between cavities can be provided with a tap-off, instead of the directional couplers used in \cite{tantawi_2006}, simplifying the system further. In this case, the cavities will be coupled and this needs to be taken into account for the design. One possible implementation of these tap-offs is shown in Figure \ref{fig:linac01}. It uses a waveguide with its narrow wall along the radial direction of the structure, which allows for a compact mechanical structure. Then a dual tap-off from each side, followed by an E-plane bend, provides the feed to two sections of the accelerator structure simultaneously through two coupling slots. Note that the feeding waveguide field has an odd symmetry around the two coupling slots, and hence it’s perfectly aligned to feed the $\pi$-mode of a standing wave accelerator structure. This way one can feed a number of accelerator structure sections with only a series of tap-offs that are half the number of sections fed; thus reducing the mechanical complexity of the overall structure. Note also that, as shown in Figure \ref{fig:linac01}, it is possible to design the overall tap-off network from identical tap-offs separated by an integer number of free space wavelength. At the same time the amplitude of the output signal at each tap off is equal and the phases of the output of the tap-offs are equal. Just as well, one could have designed the system with a $\pi$ phase shift between outputs to feed every odd number of cavities rather than even, the case shown in Figure \ref{fig:linac01}. Finally, note that the position of the short-circuit at the end of the distribution system plays a crucial role in the design and has to be chosen carefully to achieve this performance. 

The overall performance of the device allows for an extremely high repetition rate well above 10~kHz for short pulses on the order of 500~ns. With an operational gradient of  20~MV/m the maximum temperature rise for the iris  with $a/\lambda\approx 0.05$ is roughly 15~$^o$C. The peak power requirement to achieve this gradient is 1.6~MW, which leaves plenty of power from the 6~MW klystron to account for beam loading and the gun.  A summary of parameters for the RF structures is given in Table~\ref{tab:rfstructures}.

%%%%%%%%%%%%%%%%%%%%%%%%%%%%%%%%%%%%%%%%%%%%%%%%%%%%%%%%%%%%%%%%%%%%%%%%%%
\section{RF transmitter and power distribution}

A diagram of the x-band RF transmitter and microwave distribution system is shown in Figure \ref{fig:rfwaveguide} and the high-level RF transmitter specifications are given in Table~\ref{tab:rftransmitter}. A single-klystron, pulsed RF source provides adjustable peak power to the RF gun, RF deflector, and linac through the waveguide system.  Using remotely controlled couplers and phase shifters, the output power from the klystron is variably coupled to each of the three RF structures. An isolator protects the klystron by reducing reflected power in case of fill-time reflections or waveguide arcs. An optical arc detector is placed directly at the klystron RF output flange. To protect the klystron output window, this sensing system removes RF drive in a few microseconds, should visible light be detected above background either at the klystron window, or in the direction of the isolator. A high power dummy load is located at the end of the waveguide line to dissipate power not consumed by the gun, deflector, or linac.  The waveguide system is insulated with $\mathrm{SF_6}$. The RF transmitter and waveguide system operate in a lab environment and are designed to operate continuously at the maximum repetition rate of 1~kHz, 12 hours per day, 5 days per week, for 48 weeks per year.

\subsection{Klystron and modulator}

The heart of the RF transmitter is the L-3 L6145-01 klystron. This -01 version is an upgrade (from 5 to 6 MW) of the L6145 \cite{Kirshner_2009}. The cathode-pulsed klystron is fixed-tuned, electromagnetically focused, and operates at 9.3~GHz with a 30~MHz, 1~dB instantaneous bandwidth.  A 30 W solid-state power amplifier (SSPA) provides input power. The klystron outputs 6~MW peak RF power, 500~ns pulse widths, at up to 1000 pulses per second.  The klystron can operate at significantly higher than the 6~kW RF average power and 500~ns pulse widths we specified. L-3 has tested to 10.5~kW, 2.5~$\mu$s, and 700~Hz. An isolated collector allows for beam transmission monitoring. The tube is very compact at 74 cm length and 41 kg weight.  For RF stability reasons, the klystron body may be cooled with temperature-regulated water.  Test data indicates the klystron saturates at 6.0 MW output for 16 W input, for a gain of 61.5 dB.

\begin{figure}[t]
\includegraphics[width=0.45\textwidth]{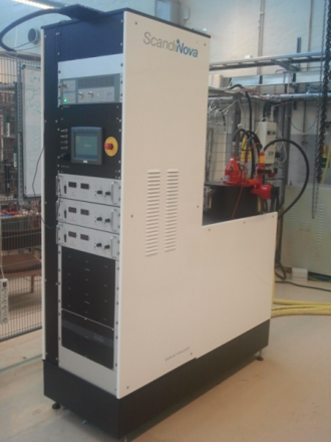}
\caption{Photograph of the compact RF transmitter that provides all power to photoinjector and linac.}
\label{fig:rftransmitterpic}
\end{figure}

\begin{table}[h]
\caption{RF transmitter specifications. Stability values are pulse-to-pulse variation over a 5 minute duration. \label{tab:rftransmitter} }
\begin{center}
\begin{tabular}{l c c}\hline\hline\
Parameter & Value & Unit \\
\hline
RF peak power & 6.0 & MW  \\
RF frequency & $9300\pm 15$ &  MHz\\
Peak-to-peak phase stability &  $\pm 0.15$ & RF deg.  \\
Peak-to-peak amp stability &  $\pm 0.05$ & \%  \\
RF pulse width & 500 & ns \\
RF average power & 3.0 & kW \\
Repetition rate & 1000 & Hz \\
\hline \hline
\end{tabular}
\end{center}
\vspace{-3mm}
\end{table}

\begin{table}[h]
\caption{Klystron modulator basic specifications. Weight includes klystron, focus coil, and focus coil power supplies.}
\label{tab:modulator} 
\begin{center}
\begin{tabular}{l c c}\hline\hline\
Parameter & Value & Unit \\
\hline
Peak output voltage & 143 & kV  \\
Peak output current & 98 &  A\\
Klystron perveance &  1.8 & micropervs  \\
Rise time &  600  & ns  \\
Flattop & 700 & ns \\
Repetition rate & 1000 & Hz \\
Average output power & 25 & kW \\
Size & $1.6\times0.7\times2.1$ & m \\
Weight & 910 & kg \\
\hline \hline
\end{tabular}
\end{center}
\vspace{-3mm}
\end{table}

\begin{table}[h]
\caption{Sensitivity (pushing factors) of RF phase and amplitude to klystron and solenoid parameters at saturation.}
\label{tab:rfpushing} 
\begin{center}
\begin{tabular}{l c c}\hline\hline\
Parameter & Phase & Amplitude \\
\hline
RF input power & 0.20 deg/W & 0.0 dB/dB \\
Beam voltage & 7.43 deg/kV & 154 kW/kV\\
Focus coil current & 0.35  deg/\% & 141 kW/\% \\
Filament current & 0.05  deg/A & 3.1 kW/A \\
Body temperature &  0.36 deg/\degree C  & 1.1 kW/\degree C \\
\hline \hline
\end{tabular}
\end{center}
\vspace{-3mm}
\end{table}

\begin{table}[h]
\caption{Klystron modulator stability specifications.  Pulse parameters are for klystron beam voltage or current over a 5 minute interval.}
\label{tab:modulatorstability} 
\begin{center}
\begin{tabular}{l c c}\hline\hline\
Parameter & Value & Unit \\
\hline
Pulse-to-pulse stability & $\pm100$ & PPM peak  \\
Leading edge timing &$\pm5$  &  ns peak\\
Flat top ripple &  $\pm250$ & PPM peak  \\
Flat top droop &  $\pm1000$  & PPM peak  \\
Filament current & $\pm400$ & PPM peak \\
Transformer reset current & $\pm1000$ & PPM peak \\
Focus coil current & $\pm100$ & PPM peak \\
\hline \hline
\end{tabular}
\end{center}
\vspace{-3mm}
\end{table}

The klystron cathode modulator (Figure \ref{fig:rftransmitterpic}) is model K1-P manufactured by ScandiNova Systems AB \cite{Hartman_2011,Elmquist_2011}. Their IGBT-switched, fractional-turn pulse transformer topology provides a flat (low droop and straight-line deviation) and highly stable beam voltage for klystrons. The pulse stability is much improved over older technologies based on pulse-forming networks.  Table \ref{tab:modulator} gives the klystron modulator basic specifications. We required that the modulator provide the peak beam voltage necessary to produce 6~MW RF power plus 10\% at the maximum repetition rate. 

The klystron modulator is made up of a 19 in. equipment rack and oil tank, mounted together on a common chassis with oil pan. The modulator DC power supplies and switching units, control system, and local control touch screen are mounted in the equipment rack. The oil tank houses the pulse transformer, droop compensation circuitry, and core reset isolation inductor, and supports the klystron and focus coil on its lid.

\subsection{RF stability}

The greatest technical challenge and most important design objective of the transmitter equipment is the RF stability requirement. Over a five minute period, RF pulse-to-pulse phase and amplitude stability are to be within $\pm0.15$ degrees and $\pm0.05\%$, respectively. For either phase or amplitude, we further define the stability specification to be the peak deviations (plus or minus) from a target value.  

To gain understanding of what control is required to meet these stability specifications one must know what parameters affect the RF output pulse and the sensitivities for each of them. The klystron vendor has supplied modeling data \cite{Kowalczyk_2012} using the same software (NRL/SAIC code TESLA, L-3 code DEMEOS, and Ansoft code Maxwell) used to design the klystron.  For the case of saturated operation, Table \ref{tab:rfpushing} shows the sensitivity of RF phase and amplitude output to various parameters.

Table \ref{tab:modulatorstability} compiles specifications important to RF intra- and inter-pulse stability for the klystron beam voltage and ancillary power supply currents. These were determined through the phase and amplitude pushing factors and by determining what has been demonstrated by the klystron modulator vendors.   Short term stability, say over ten pulses, will span 10~ms and should be dominated by high frequency power supply output variations. As the time span increases, power supply line regulation and cooling are important.  We note that 5 PPM control of DC power supplies has been achieved \cite{Tian_magnetps} with add-on controllers.

%%%%%%%%%%%%%%%%%%%%%%%%%%%%%%%%%%%%%%%%%%%%%%%%%%%%%%%%%%%%%%%%%%%%%%%%%%
\section{Summary}

There is a vast gulf between the x-ray performance of a modern synchrotron beamline and a home-laboratory source based on bremsstrahlung.  The synchrotron is 12 orders of magnitude brighter and produces several orders of magnitude more flux, enabling qualitatively different science.  However advances in laser and accelerator technologies have shrunk the size and cost of an accelerator-based x-ray source to a scale suitable for the home lab, bringing the power of synchrotron-like radiation into the home lab.  Our design estimates of x-ray performance include flux of $5 \times 10^{11}$~photons/sec in a 5\% bandwidth and brilliance of $2 \times 10^{12}$ photons/(sec mm$^2$ mrad$^2$ 0.1\%) and RMS pulse duration of 0.5 ps.  The proliferation of such sources is likely to have a large impact on x-ray science and medicine, allowing a wide and diverse range of researchers to probe atomic-scale structure, high resolution medical imaging, and femtosecond-scale time dynamics without the hurdles of obtaining beamtime at a remote facility.  The impact of introducing students and researchers from academia, medicine, and industry to powerful x-ray techniques is likely to expand the depth and diversity of science and scientists at the major facilities as well as the x-ray community grows in ways that it cannot today.

The new technologies include highly stable diode-pumped table-top lasers with high peak and average power, and very compact high efficiency accelerators wtih modest power requirements.  The brilliance of the compact source is several orders of magnitude higher than today's lab sources and its flux is substantially higher, while it is capable of producing tunable, polarized monochromatic x-rays.  Enhancements to the proposed source such as coherent emission and operation with superconducting accelerators and coherently pumped laser cavities can yield large increases in performance with continuing R\&D.  With regard to pulse length and source size, the compact source outperforms the large synchrotrons.

We have described the accelerator that relies on a high efficiency copper linac and photoinjector powered by a single small klystron and solid-state modulator to produce stable, high-brightness electron beams at energies of a few tens of MeV.  The high efficiency allows the linac to pulse at 1 kHz with a train of 100 electron bunches in each pulse for an effective repetition rate of 100 kHz, many orders of magnitude beyond standard copper linac performance.

The laser uses cryo-cooled ytterbium crystals to produce the required high average power and short picosecond pulses with exceptional beam quality.  The laser pulse is coupled into a ringdown cavity tuned to the electron bunch train spacing so that the laser pulse repeatedly collides with different bunches.  The collision of the electron and laser beams produces bright x-rays via inverse Compton scattering, which requires both beams to be focused to a micron-sized spot and to maintain picosecond or shorter length in order to produce substantial x-ray flux.  Example x-ray beamlines using nested Kirkpatrick-Baez optics have been described that are optimized for the ICS beam properties of small source size and large divergence.

%%%%%%%%%%%%%%%%%%%%%%%%%%%%%%%%%%%%%%%%%%%%%%%%%%%%%%%%%%%%%%%%%%%%%%%%%%
\section{Acknowledgements}
This work was supported by DARPA grant N66001-11-1-4192, NSF grant DMR-1042342, and DOE grants DE-FG02-10ER46745 and DE-FG02-08ER41532, and the Center for Free-Electron Laser Science through the DESY-MIT Collaboration.


\begin{thebibliography}{1}
\bibitem{roentgen_1896} W.C. R{\"o}ntgen, ``On a new type of rays,”  Nature 53, 274-276 (1896).
\bibitem{graves_2012} W.S. Graves, P. Piot, F.X. K{\"a}rtner, and D.E. Moncton, ``Intense Superradiant X Rays from a Compact Source Using a Nanocathode Array and Emittance Exchange,” Phys Rev Lett 108, 263904 (2012).
\bibitem{graves_2013} W.S. Graves, K.K. Berggren, J. Bessuille, P. Brown, S. Carbajo, J. Derksen, A. Fallahi, R. Hobbs, K.-H. Hong, W. R. Huang, E. Ihloff, F.X. K{\"a}rtner, P. D. Keathley, D. Mihalcea, D.E. Moncton, E. Nanni, Ph. Piot, K. Ravi, F.Scheiba, M. Swanwick, L. F. Velásquez-García, I. Viti, L.J. Wong, X.Wu, Y. Yang, L. Zapata, and Y. Zhou, 35th International Free Electron Laser Conference, pp. 757-761 (2013) New York, NY
\bibitem{brown_2004a} W.J. Brown, S.G. Anderson, C.P.J. Barty, S.M. Betts, R. Booth, J.K. Crane, R.R. Cross, D.N. Fittinghoff, D.J. Gibson, F.V. Hartemann, E.P. Hartouni, J. Kuba, G.P. Le Sage, D.R. Slaughter, A.M. Tremaine, A.J. Wooton, P.T. Springer, and J.B. Rosenzweig,  ``Experimental characterization of an ultrafast Thomson scattering x-ray
source with three-dimensional time and frequency-domain analysis", Phys. Rev. ST-AB 7, 060702 (2004)
\bibitem{xu_2014} H.S. Xu, W.H. Huang, C.X. Tang, and S.Y. Lee, Phys. Rev. ST-AB 17, 070101 (2014)
\bibitem{rigaku_frx} http://www.rigaku.com/products/protein/frx
\bibitem{krafft_2004} G.A. Krafft, ``Spectral Distributions of Thomson-Scattered Photons from High-Intensity Pulsed Lasers", Phys. Rev. Lett. 92, 204802 (2004)
\bibitem{hartemann_2005b} F.V. Hartemann, W. J. Brown, D. J. Gibson, S. G. Anderson, A. M. Tremaine, P. T. Springer, A. J. Wootton,
E. P. Hartouni, and C. P. J. Barty, ``High-energy scaling of Compton scattering light sources", Phys. Rev. ST-AB 8, 100702 (2005)
\bibitem{albert_2011} F. Albert, S. G. Anderson, D. J. Gibson, R. A. Marsh, C. W. Siders, C. P. J. Barty, and F. V. Hartemann, "Three-dimensional theory of weakly nonlinear Compton scattering", Phys. of Plasmas 18, 013108 (2011)
\bibitem{ride_1995} S.K. Ride, E. Esarey, and M. Baine, "Thomson scattering of intense lasers from electron beams at arbitrary interaction angles", Phys. Rev. E 52, 5425 - 5442  (1995) 
\bibitem{kim} see e.g. K.-J. Kim,``Characteristics of Synchrotron Radiation", section 2 of X-ray Data Booklet, http://xdb.lbl.gov/
\bibitem{parmela} L. Young PARMELA Reference Manual, Los Alamos report LA-UR-96-1835
\bibitem{brown_2004b} W.J. Brown and F.V. Hartemann, ``Three-dimensional time and frequency-domain theory of femtosecond
x-ray pulse generation through Thomson scattering", Phys. Rev. ST-AB 7, 060703 (2004)
\bibitem{brown_2004c} W.J. Brown and F.V. Hartemann, ``Brightness Optimization of Ultra-Fast Thomson Scattering X-ray Sources", 11th Advanced Accelerator Concepts Workshop. AIP Conference Proceedings, Volume 737, pp. 839-845 (2004)
\bibitem{settens} Settens, Charles, et al. ``Critical dimension small angle X-ray scattering measurements of FinFET and 3D memory structures.'' SPIE Advanced Lithography. International Society for Optics and Photonics, 2013.
\bibitem{luiten} O. J. Luiten, S. B. van der Geer, M. J. de Loos, F. B. Kiewiet, and M. J. van der Wiel, Phys. Rev. Lett 93, 094802 (2004)
\bibitem{ferrario} M. Ferrario, J. E. Clendenin, D. T. Palmer, J. B. Rosenzweig, and L. Serafini, ``HOMDYN study for the LCLS RF photo-injector'', SLAC-PUB-8400 (2000)
\bibitem{dowell_2009} D.H. Dowell and J.F. Schmerge,  ``Quantum efficiency and thermal emittance of metal photocathodes", Phys. Rev. ST-AB 12, 074201 (2009)
\bibitem{loos} Loos, Henrik. ``LCLS accelerator operation and measurement of electron beam parameters relevant for the x-ray beam.'' SPIE Optics+ Optoelectronics. International Society for Optics and Photonics, 2013.
\bibitem{Fan} T. Y. Fan, D. J. Ripin, R. L. Aggarwal, J. R. Ochoa, B. Chann, M. Tilleman, and J. Spitzberg, ``Cryogenic Yb3+-Doped Solid-Stae Lasers,” IEEE J. Sel. Top. Quantum Electron. 13, 448 (2007).
\bibitem{Hong_OL2} K.-H. Hong, J. Gopinath, D. Rand, A. Siddiqui, S.-W. Huang, E. Li, B. Eggleton, John D. Hybl, T. Y. Fan and F. X. K{\"a}rtner, ``High-energy, kHz-repetition-rate, ps cryogenic Yb:YAG chirped-pulse amplifier,” Opt. Lett. 35, 1752 (2010).
\bibitem{Akahane} Y. Akahane, M. Aoyama, K. Ogawa, K. Tsuji, S. Tokita, J. Kawanaka, H. Nishioka, and K. Yamakawa, ``High-energy, diode-pumped, picosecond Yb:YAG chirped-pulse regenerative amplifier for pumping optical parametric chirped-pulse amplification,” Opt. Lett. 32, 1899 (2007). 
\bibitem{Metzger} T. Metzger, A. Schwarz, C. Y. Teisset, D. Sutter, A. Killi, R. Kienberger, and F. Krausz, ``High-repetition-rate picosecond pump laser based on a Yb:YAG disk amplifier for optical parametric amplification,” Opt. Lett. 34, 2123-2125 (2009). 
\bibitem{zapata_2013} Luis E. Zapata, Hua Lin, Huseyin Cankaya, Anne-Laure Calendron, Wenqian Huang, Kyung-Han Hong and Franz X. K{\"a}rtner, ``Cryogenic Composite Thin Disk High Energy Pulsed, High Average Power, Diffraction Limited Multi-Pass Amplifier'', ASSL (Paris), 2013.
\bibitem{hong_2014} K.-H. Hong, C.-J. Lai, J. Siqueira, P. Krogen, J. Moses, C.-L. Chang, G. J. Stein, L. E. Zapata, and F. X. K{\''a}rtner, “Multi-mJ, kHz, 2.1-$\mu$m optical parametric chirped pulse amplifier and high-flux soft X-ray high-harmonic generation,” Optics Letters 39, 3145-3148 (2014)
\bibitem{Fu_JOSAB} Xing Fu, Kyung-Han Hong, Li-Jin Chen, and Franz X. K{\"a}rtner, ``Performance scaling of high-power picosecond cryogenically-cooled Yb:YAG multipass amplification,” J. Opt. Soc. Am. B 30, 2798-2808 (2013).
\bibitem{strickland_1985} D. Strickland and G. Mourou, Opt. Commun. 56, 219 (1985).
\bibitem{hong_2008} K. H. Hong, A. Siddiqui, J. Moses, J. Gopinath, J. Hybl, F. O. Ilday, T.Y. Fan and F. X. K{\"a}rtner, "Generation of 287-W, 5.5-ps pulses at 78-MHz repetition rate from a cryogenically-cooled Yb:YAG amplifier seeded by a fiber chirped-pulse amplification system”, Opt. Lett. 33, 2473-2475 (2008)
\bibitem{ripen_2005} D. J. Ripin, J.R. Ochoa, R.L. Aggarwal, and T.Y. Fan, ``300-W Cryogenically Cooled Yb:YAG Laser”, IEEE JQE 41, 1274 (2005)
\bibitem{jovanovic} I. Jovanovic, M. Shverdin, D. Gibson, and C. Brown, ``High-power pulse recirculation for inverse Compton scattering-produced $\gamma$-rays" Nuc. Inst. Meth. A 578, 160-171 (2007)
\bibitem{Newport_HR} ``High Performance SuperMirrors.'' Newport, n.d. Web. 24 March 2014.
\bibitem{qtf_AR} ``Thin Film Coatings.'' Quality Thin Films, Inc., n.d. Web. 24 March 2014
\bibitem{castech_lbo} ``CASTECH - Crystal Catalog.'' Castech, Inc., 2013. Web. 24 March 2014.
\bibitem{hong_2009} Hong, Kyung-Han, et al. ``130-W picosecond green laser based on a frequency-doubled hybrid cryogenic Yb: YAG amplifier." Optics express 17.19 (2009): 16911-16919
\bibitem{zhang_1998} Zhang, Jing-yuan, et al. ``Second-harmonic generation from regeneratively amplified femtosecond laser pulses in BBO and LBO crystals." JOSA B 15.1 (1998): 200-209
\bibitem{hunt_1989} J.T. Hunt, D.R. Speck: Opt. Eng. 28, 461 (1989)
\bibitem{hunt_1993} J.T. Hunt, K.R. Manes, P.A. Renard: Appl. Opt. 32, 5973 (1993)
\bibitem{gibson_2010} D. J. Gibson, F. Albert, S. G. Anderson, S. M. Betts, M. J. Messerly, H. H. Phan, V. A. Semenov, M. Y. Shverdin, A. M. Tremaine, F. V. Hartemann, C. W. Siders, D. P. McNabb, and C. P. J. Barty, ``Design and operation of a tunable MeV-level Compton-scattering-based $\gamma$-ray source'', Phys. Rev. ST-AB 13, 070703 (2010)
\bibitem{albert_2011b} F. Albert, S. G. Anderson, D. J. Gibson, R. A. Marsh, S. S. Wu, C. W. Siders, C. P. J. Barty, and F. V. Hartemann, ``Design of narrow-band Compton scattering sources for nuclear resonance fluorescence'', Phys. Rev. ST-AB 14, 050703 (2011)
\bibitem{nanni_2014} E.A. Nanni, W.S. Graves, and P. Piot, ``Nanometer scale coherent current modulation via a nanotip cathode array and emittance exchange'', Proceedings of 2014 Int'l Particle Accelerator Conf. 1952-1955 (2014) Dresden, Germany
\bibitem{kim_rfgun} K.-J. Kim, ``RF and space-charge effects in laser-driven RF electron guns,'' Nucl. Inst. Meth. A 275, 201-218 (1989)
\bibitem{Haimson_1999} J. Haimson,  B. Mecklenburg, and G. Stowell, ``A field symmetrized dual feed 2 MeV RF gun for a 17 GHz electron linear accelerator'', proceedings of the eighth workshop on advanced accelerator concepts, 653-667, Baltimore, MD (1998)
\bibitem{Nantista_2004} C. Nantista, S. Tantawi, and V. Dolgashev, ``Low-field accelerator structure couplers and design techniques'', Phys. Rev. ST-AB 7, 072001 (2004)
\bibitem{tantawi_TBB} S. Tantawi et. al., ``The optimization of linear accelerator structures with distributed coupling,''  to be published.
\bibitem{tantawi_patent2014} Sami G. Tantawi and Jeffrey Neilson, ``Distributed Coupling High Efficiency Linear Accelerator'', US Patent Application 61/674262,  Patent Pending.
\bibitem{tantawi_2006} Sami G. Tantawi, ``rf distribution system for a set of standing-wave accelerator structures,'' Phys. Rev. ST-AB 9, 112001 (2006).
\bibitem{schaffer_1965} G. Schaffer, ``High Power UHF Components for DESY'', IEEE Trans. On Nucl. Sci. NS-12, No. 3, 208 (1965)
\bibitem{sundelin_1977} R. M. Sundelin, J. L. Kirchgessner, and M. Tigner, ``Parallel Coupled Structure,'' IEEE Trans. on Nucl. Sci. NS-24, No. 3, 1686-1688 (1977)
\bibitem{yu_2010} Yu. Chernousov, V. Ivannicov, I. Shebolaev, A. Levichev, and V. Pavlov, ``Characteristics of Parallel Coupled Accelerating Structure,'' Proceedings of IPAC'10, Kyoto, Japan, 2010, 3765-3767
\bibitem{Kirshner_2009} M.F. Kirshner, R.D. Kowalczyk, C.B. Wilsen, R.B. True, I.A. Chernyavskiy, and A. Vlasov, ``High power X-band klystron", Vacuum Electronics Conference (IVEC), 2009 IEEE International, pp.535-536, April, 2009
\bibitem{Hartman_2011} M. Lindholm, W. Crewson, and K. Elmquist, ``Pulse to Pulse Stability at Parts per Million (ppm) 
Level,” Proceedings of 18th IEEE International Pulsed Power Conference, June 2011 pp. 1593 - 1597, (Chicago, IL)
\bibitem{Elmquist_2011} K. Elmquist, M. Lindholm, and W. Crewson, ``High Power Pulse Quality Using Solid State 
Technology”, Proceedings of 18th IEEE International Pulsed Power Conference, June 2011 pp. 1598-1601, (Chicago, IL)
\bibitem{Kowalczyk_2012} R. Kowalczyk, ``Predicted Amplitude and Phase Pushing Factors on the L6145 X-Band Klystron,” private communication, L-3 Communications Electron Devices, September 2012
\bibitem{Tian_magnetps} Y. Tian, W. Louie, J. Ricciardelli, L.R. Dalesio, and G. Ganetis, ``Power supply control system of NSLS-II,” Proceedings of ICALEPCS2009, 385-387 (2009), Kobe, Japan 

\end{thebibliography}
\end{document}